\documentclass{article}

\usepackage{microtype}
\usepackage{graphicx}
\usepackage{subcaption}
\usepackage{booktabs} %
\usepackage{makecell}
\usepackage{enumitem}
\usepackage{placeins}

\usepackage{hyperref}
\usepackage{multirow}
\usepackage{pifont}
\usepackage{xcolor}

\newcommand{\CHECK}{\textcolor{green}{\ding{52}}} %
\newcommand{\CROSS}{\textcolor{red}{\ding{55}}} %
\newcommand{\PART}{\textcolor{orange}{\ding{52}\rotatebox[origin=c]{-9.2}{\kern-0.7em\ding{55}}}}

\usepackage[accepted]{icml2026}

\usepackage{amsmath}
\usepackage{amssymb}
\usepackage{mathtools}
\usepackage{amsthm}
\usepackage{float}

\usepackage[capitalize, noabbrev]{cleveref}
\theoremstyle{plain}

\theoremstyle{definition}

\theoremstyle{remark}

\usepackage[textsize=tiny]{todonotes}
\usepackage{multirow}
\icmltitlerunning{CMI-RewardBench: Evaluating Music Reward Models with Compositional Multimodal Instruction}

\begin{document}
\twocolumn[ \icmltitle{CMI-RewardBench: Evaluating Music Reward Models\\ with Compositional Multimodal Instruction }

  \icmlsetsymbol{equal}{*}

  \begin{icmlauthorlist}
    \icmlauthor{Yinghao Ma}{qmul,equal} \icmlauthor{Haiwen Xia}{pku,equal} \icmlauthor{Hewei Gao}{tum,dtu}
    \icmlauthor{Weixiong Chen}{qmul} \icmlauthor{Yuxin Ye}{you} \icmlauthor{Yuchen Yang}{soochow}
    \icmlauthor{Sungkyun Chang}{qmul}
    \icmlauthor{Mingshuo Ding}{pku} \icmlauthor{Yizhi Li}{man} \icmlauthor{Ruibin Yuan}{hk}
    \icmlauthor{Simon Dixon}{qmul} \icmlauthor{Emmanouil Benetos}{qmul}

  \end{icmlauthorlist}

  \icmlaffiliation{qmul}{Queen Mary University of London, London, UK}
  \icmlaffiliation{pku}{Peking University, Beijing, China} \icmlaffiliation{tum}{Technical University of Munich, Munich, Germany}
  \icmlaffiliation{dtu}{Technical University of Denmark, Lyngby, Denmark}
  \icmlaffiliation{you}{Beijing University of Post and Telecommunications, Beijing, China}
  \icmlaffiliation{soochow}{Soochow University, Suzhou, China} \icmlaffiliation{man}{University of Manchester, Manchester, UK}
  \icmlaffiliation{hk}{Hong Kong University of Science and Technology, Hong Kong, China}

  \icmlcorrespondingauthor{Yinghao Ma}{yinghao.ma@qmul.ac.uk} \icmlcorrespondingauthor{Emmanouil Benetos}{emmanouil.benetos@qmul.ac.uk}

  \icmlkeywords{music reward modeling, AI music evaluation, compositional multimodal instruction, music generation}

  \vskip 0.3in ]

\printAffiliationsAndNotice{\icmlEqualContribution} %

\begin{abstract}
  While music generation models have evolved to handle complex multimodal inputs
  mixing text, lyrics, and reference audio, evaluation mechanisms have lagged
  behind.
  In this paper, we bridge this critical gap by establishing a comprehensive
  ecosystem for music reward modeling under \textbf{Compositional Multimodal
    Instruction (CMI)}, where the generated music may be conditioned on text
  descriptions, lyrics, and audio prompts.
  We first introduce \textbf{CMI-Pref-Pseudo}, a large-scale preference dataset
  comprising 110k pseudo-labeled samples, and \textbf{CMI-Pref}, a high-quality,
  human-annotated corpus tailored for fine-grained alignment tasks. To unify
  the evaluation landscape, we propose \textbf{CMI-RewardBench}, {a unified benchmark that evaluates music reward models on heterogeneous samples across musicality, text–music alignment, and compositional instruction alignment.}
  Leveraging these resources, we develop \textbf{CMI reward models (CMI-RMs)},
  a parameter-efficient reward model family capable of processing
  heterogeneous inputs. We evaluate their correlation with human judgment scores
  on musicality and alignment on CMI-Pref along with previous datasets.
  Further experiments demonstrate that CMI-RM not only correlates strongly
  with human judgments, but also enables effective \textbf{inference-time
    scaling} via top-$k$ filtering.Code is available
  at \href{https://github.com/Haiwen-Xia/CMI-RewardBench}{GitHub}. Model weights:
  \href{https://huggingface.co/HaiwenXia/CMI-RM}{CMI-RM}. Datasets:
  \href{https://huggingface.co/datasets/HaiwenXia/cmi-pref-pseudo}{CMI-Pref-Pseudo}
  and \href{https://huggingface.co/datasets/HaiwenXia/cmi-pref}{CMI-Pref}.
\end{abstract}

\section{Introduction}

The rapid advancement of Artificial Intelligence Generated Content (AIGC) has significantly
impacted the creative industries, with music generation emerging as one of the
cornerstones for numerous commercial applications in music, movie and entertainment
industries~\citep{cao2025survey, ren2025aigc, ma2024foundation}. Despite the proliferation
of sophisticated generative models, evaluating their outputs remains a
fundamental challenge. Generally speaking, music evaluation requires assessing
both musicality and instruction following. However, advancements in generative
models now necessitate this assessment under flexible multimodal conditions, such
as text-only, lyric-guided, and audio-referenced inputs.%

Developing these evaluation models is hindered by a critical data scarcity. Although
large-scale user interaction data exists in music recommendation (e.g.,
Spotify Million Playlist~\cite{papreja2019representation}), it fundamentally
captures user-item affinity—a global preference for genre styles or
playlists—rather than generative alignment, which demands assessment of perceptual
quality and precise instruction-following. Such recommendation datasets lack
the fine-grained, comparative rankings of generated samples against complex,
multimodal instructions (such as interwoven lyrics, text descriptions, and reference
audio) required to train alignment models~\cite{pam}.

\begin{figure*}
  \centering
  \includegraphics[width=.65\linewidth]{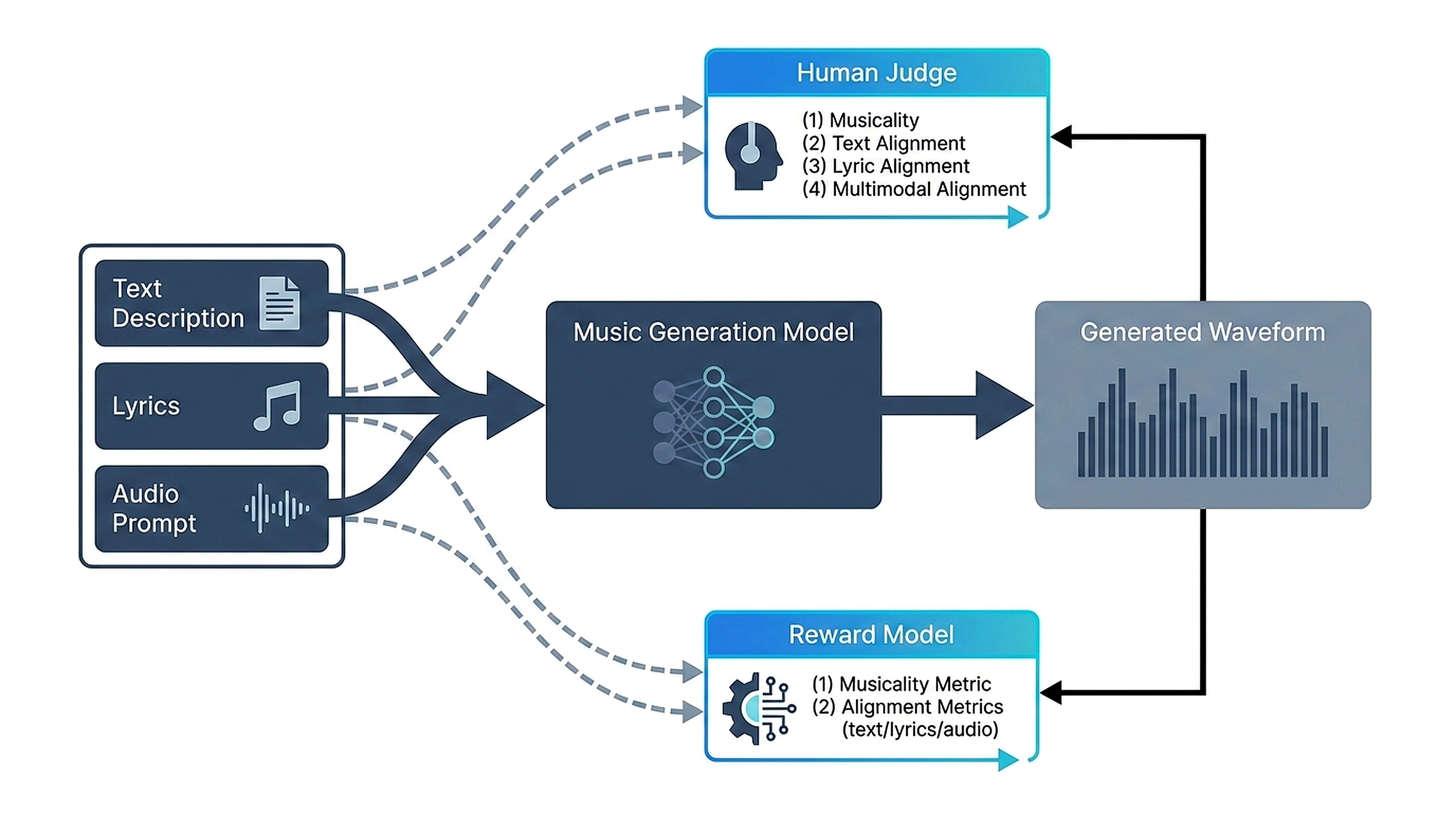}
  \caption{Reward models should act as proxies for human evaluation under
    compositional multimodal instructions (CMI). Human judges evaluate a generated
    waveform with respect to the provided prompt—text descriptions, lyrics, and/or
    reference audio—considering both musicality and instruction alignment.
    However, current reward models are typically fragmented: they either score
    musicality from audio alone or capture alignment for a single modality pair,
    and thus cannot reliably proxy human judgments across flexible CMI inputs.}
  \label{fig:teaser}
\end{figure*}

Consequently, evaluation methodologies have struggled to bridge this gap.
Traditional metrics like Fréchet Audio Distance (FAD)~\citep{kilgour2019frechet}
operate at the distribution level, failing to provide the sample-level signals
necessary for post-training or filtering. More recent approaches, such as
SongEval~\citep{songeval}, PAM~\citep{pam}, and various MOS predictors~\citep{audiobox},
have advanced the field by offering sample-level scoring. However, these
efforts remain fragmented and narrowly specialized. They typically focus on isolated
attributes (e.g., only caption alignment) and rely on rigid input assumptions,
whereas state-of-the-art music generation models already support flexible
input combinations, ranging from simple text prompts to interwoven lyrics and audio
references. This growing mismatch between model capabilities and evaluation
methodologies is highlighted in \autoref{fig:teaser}.

We argue that effective evaluation requires compositional alignment, defined here
not merely as adherence to simultaneous constraints, but as the capability of
a unified model to adaptively agree with human preferences across these
optional and varying input conditions. Specifically, the model should assign
scores or rankings that consistently reflect human judgments regarding both
musical quality and instruction adherence, regardless of whether inputs are text-only,
lyric-guided, or audio-referenced. A framework capable of judging this
versatility is currently missing.

To bridge this gap, our proposed \textbf{CMI-RewardBench} integrates diverse task-specific
datasets to vigorously evaluate whether a single reward model can judge generation
quality against the heterogeneous instruction sets inherent to modern AIGC flows.

In this paper, our contributions are three-fold:
\begin{enumerate}
  \item We construct \textbf{CMI-Pref-Pseudo}, containing 110k samples labeled
        via a robust pipeline using Qwen3-Omni \cite{qwen3} with consistency filtering.
        Complementing this, we introduce \textbf{CMI-Pref}, a high-quality corpus of
        4,027 pairs annotated by 31 human experts. These annotations capture fine-grained
        preferences for musicality, alignment, and confidence levels across
        diverse genres, instruments, and multimodal prompts (including lyrics and audio-to-audio
        conditioning).

  \item We propose \textbf{CMI-RewardBench}, a unified benchmark for music reward
        models. By integrating existing resources (PAM, MusicEval, Music Arena)
        with our CMI-Pref test split, this benchmark evaluates models on five distinct
        tasks ranging from absolute musicality scoring to complex compositional alignment.
        This unified approach serves as a rigorous testbed for model \textbf{versatility}
        across optional input settings. Our baseline evaluations on this benchmark
        expose a significant capability gap, revealing that even state-of-the-art multimodal
        LLMs (e.g., Gemini-2.5-Pro) struggle to exceed 80\% agreement with human
        preferences.

  \item We develop \textbf{CMI-RM}, a family of music reward models supporting
        compositional conditioning over text, lyrics, and audio. Uniquely
        supporting all evaluation settings in CMI-RewardBench via a single, parameter-efficient
        architecture ($\sim$30M), CMI-RM achieves performance comparable to or better
        than specialized open-source baselines like SongEval. Furthermore, we demonstrate
        that CMI-RM provides measurable benefits when used for top-k filtering,
        enabling ``inference-time scaling'' for music generation. %
\end{enumerate}

\section{Related Work}

\subsection{RLHF for LLMs and MLLMs}
Reinforcement Learning from Human Feedback (RLHF) has successfully aligned generative
models with human intent by replacing heuristic proxies (e.g., ROUGE) with learned
reward models \citep{stiennon2020learning, ouyang2022training, bai2022training}.
This paradigm has successfully extended to multimodal domains, including text-to-image
generation \citep{kirstain2023pick, xu2023imagereward} and video synthesis \citep{DBLP:conf/acl/AhnCYKC24, liu2025improving},
to address aesthetic quality and semantic alignment. Recent research has also
applied preference optimization to speech synthesis to improve naturalness and
bridge inference-time distribution gaps \citep{zhang2024speechalign, zhang2025speechjudge}.
Concurrently, the use of LLMs as scalable evaluators \citep{zheng2023judging, gu2024survey}
has emerged as a robust alternative to human experts.
While these works address text, vision, and speech, music evaluation remains
under-explored and fragmented. Current music metrics lack the capacity to judge
condition on compositional instructions.

\subsection{Evaluation Metrics of Music}

Music evaluation typically bifurcated into distribution-level quality assessment
and sample-level alignment metrics. However,
existing approaches struggle to address the complexity of compositional instructions.

\textbf{Quality and Subjective Metrics.} Distributional metrics like the
Fréchet Audio Distance (FAD) \citep{kilgour2019frechet} are the standard for global
corpus assessment, with recent variants like MAD \citep{huang2025aligning} and
KAD \citep{kad} improving correlation with human perception. For sample-level musicality,
MOS predictors such as PAM \citep{pam}, Audiobox \citep{audiobox} and SongEval
\citep{songeval} evaluate aesthetic quality. While effective for general audio,
high-performing systems like MusicRL \citep{cideron2024musicrl}, WhisQ
\cite{emon2025whisq}, QAMRO \cite{wang2025qamro}, and DRAGON \citep{bai2025dragon}
remain closed-source.

\textbf{Alignment and LLM Judging.} Alignment is primarily measured via
contrastive scores like CLAP \citep{wu2023large}, CLaMP3 \citep{wu2025clamp3universalmusic},
and MuQ-Mulan \cite{zhu2025muqselfsupervisedmusicrepresentation}. While music-specific
checkpoints improve human-preference alignment \citep{grotschla2025benchmarking},
these metrics are largely restricted to text-to-audio pairs and neglect lyrics
or audio prompts. Besides, emerging ``LLM-as-a-judge'' frameworks like AutoMV \citep{tang2025automv}
and music recommendation AutoRaters \citep{chen2024comparing} offer scalable,
multimodal evaluation with complex instructions. However, these rely on
proprietary models and lack an open-source framework for evaluating
compositional multimodal music instructions with lyrics and audio prompts
information.

\subsection{Preference Datasets and Platforms}

The development of robust reward models relies on standardized preference datasets
and evaluation platforms. Early efforts such as {MusicEval} \citep{liu2025musiceval},
\textbf{SongEval} \citep{songeval}, and {AudioEval} \citep{wang2025audioeval}
provide expert-annotated corpora for absolute quality prediction and community
benchmarks \citep{huang2025audiomos, ma2026icassp, zhang2025aesthetics}. Recent
pairwise datasets such as {AIME} \citep{grotschla2025benchmarking} and MusicPref
\citep{huang2025aligning} benchmark text-to-music systems via human comparisons,
but remain limited to text-to-music generation.

Inspired by the crowdsourcing success of {Chatbot Arena} \citep{chiang2024chatbot},
{GenAI Arena} \citep{jiang2024genai}, and {Copilot Arena} \citep{chicopilot}
etc., the recently introduced {Music Arena} \citep{kim2025music} provides a live
platform for comparative text-to-music evaluation. While these resources represent
progress, they primarily focus on text-to-music alignment. \textbf{CMI-Pref} fills
this gap with large-scale preferences for {compositional} instructions, including
lyrics and audio-to-audio conditioning.

\section{Method}

\subsection{Data Sources for Training and Benchmarking}
We use CMI-Pref-Pseudo for Bradley--Terry pre-training, the training split for CMI-Pref and
MusicEval for fine-tuning, and the test split of CMI-Pref, MusicEval along with the full PAM and Music Arena datasets for benchmarking.

\textbf{PAM.} We include 500 audio clips from the music subset. Each clip is
associated with MOS annotations for musicality and text-music alignment with a
text description.

\textbf{MusicEval.} We utilize the training and test splits,which provide
expert-validated MOS for musicality (musical impression). Due to mismatches among
audio files, filenames, and text prompts, we omit its MOS for text-music alignment.

\textbf{Music Arena.} We process 2,800 historical interaction logs from the
Arena platform up to Dec 2025. To ensure high-quality preference pairs, we remove
failed generations and filter out ``tie'' or ``both bad'' labels, since users may
apply different tolerance margins when two audios are similar. This yields 1,340
clean preference pairs. We categorize Music Arena labels as \textsc{musicality};
further discussion is provided in Appendix~\ref{Append:Musicality}.

We compare additional resources in \autoref{tab:data-statis},
including AIME, MusicPref, and SongEval, but do not include them as main benchmark
test sets: MusicPref follows a protocol similar to Music Arena, while Music Arena
is more up-to-date and contains more in-the-wild prompts from a live platform;
AIME provides two preference dimensions but is limited to text-to-music generation
with label-composed prompts; and SongEval does not provide the generation prompts
needed for prompt-alignment evaluation. These datasets lack official splits aligned
with our protocol, and future work could explore them as additional training sets.
\begin{table*}
  [t]
  \caption{Statistics of CMI-Pref and CMI-Pref-Pseudo compared with previous
    datasets. Samples here refer to pairs for preference datasets and individual audio clips for MOS datasets.}
  \label{tab:data-statis}
  \begin{center}
    \begin{small}
      \resizebox{\textwidth}{!}{
        \begin{sc}
          \begin{tabular}{lcccccccc}
            \toprule Data set                    & PAM    & MusicEval   & SongEval    & AIME   & MusicPref   & Music Arena & CMI-Pref-Pseudo & CMI-Pref \\
            \midrule Text Condition              & \CHECK & unavailable & \CROSS      & \CHECK & \CHECK      & \CHECK      & \CHECK          & \CHECK   \\
            Lyrics Condition                     & \CROSS & \CROSS      & unavailable & \CROSS & \CROSS      & \CHECK      & \CHECK          & \CHECK   \\
            Audio Condition                      & \CROSS & \CROSS      & \CROSS      & \CROSS & \CROSS      & \CROSS      & \CHECK          & \CHECK   \\
            \midrule \#samples                   & 500    & 2748        & 2,399       & 15,600 & 2,520       & 2,800       & 110k            & 4,027    \\
            \#Testing samples                    & 500    & 413         & -           & -      & -           & 1,340       & -               & 500      \\
            \#Duration of audio(hours)           & 0.83   & 16.62       & 140.54      & 16.67  & unavailable & 88.30       & 808.85          & 133.80   \\
            \#Duration of reference audio(hours) & -      & -           & -           & -      & -           & -           & 178.36          & 48.56    \\
            \#Unique Prompts                     & 100    & 384         & -           & 500    & 2,617       & 883         & 10,213          & 2,632    \\
            \#Models \& APIs                     & 5      & 31          & 7           & 12     & 7           & 17          & 23              & 23       \\
            Official split                       & full   & \CHECK      & \CROSS      & \CROSS & \CROSS      & \CROSS      & -               & \CHECK   \\
            \bottomrule
          \end{tabular}
        \end{sc}
      }
    \end{small}
  \end{center}
  \vskip -0.1in
\end{table*}

\subsubsection{CMI-Pref and CMI-Pref-Pseudo: Large-Scale Compositional Multimodal
  Preference Datasets}

\textbf{CMI-Pref} captures human preferences when music is conditioned on CMIs,
including text descriptions, lyrics, and audio prompts. \textbf{CMI-Pref-Pseudo}
provides large-scale LLM-judge labels for the same setting.

\paragraph{Data Collection}

We distilled audio from a diverse set of 12 models and 11 commercial APIs to
ensure a broad distribution of quality and style. For commercial APIs or products,
we generated samples using Suno (v3.5, v4, v4.5, v4.5-plus, v5)\footnote{\url{https://suno.com/home}},
Stable Audio 2.0\footnote{\url{https://platform.stability.ai/}}, Minimax-Music-2.0\footnote{\url{https://platform.minimaxi.com/}},
Mureka (v7.5, o2)\footnote{\url{ https://platform.mureka.ai/}}, and Loudly\footnote{\url{https://www.loudly.com/}}.
These include equal splits of instrumental and vocal tracks (with lyrics) if the
model supports lyrics as input, and equal splits of input with and without
audio prompts if applicable.

For open-source models, we generated audio from MusicGen \citep{copet2023simple},
Stable Audio Open \citep{DBLP:conf/icassp/EvansPCZTP25}, YUE \citep{DBLP:journals/corr/abs-2503-08638},
SongGen \citep{liusonggen}, AudioLDM \citep{liu2023audioldm}, AudioLDM 2 \citep{liu2024audioldm},
DiffRhythm \citep{ning2025diffrhythm}, Levo \citep{lei2025levo}, Magenta Lyria-RealTime
\citep{team2025live}, Jamify \citep{liu2025jam}, MusicLDM\citep{chen2024musicldm},
and ACE-step \citep{gong2025ace}.
35.6\% of samples
are conditioned on audio prompts for style transfer or continuation in addition
to text and lyrics. Audio caption of the audio prompt is provided by Qwen3-Omni
as additional text condition if model input cannot support audio prompt.

\paragraph{Annotation and Statistics}

The CMI-Pref-Pseudo dataset initially includes 130k samples, while retaining 110k
pairs after consistency checks (See \autoref{app:pseudo-label-generation}), spanning 47,546 generations (797.34h). It is
generated by prompting Qwen3-Omni with the prompts in \autoref{app:prompts},
yielding two-dimensional preference pairs for musicality and prompt alignment.

Following the annotation protocol described in \autoref{app:human-annotation},
31 annotators constructed {CMI-Pref}. For each pair, annotators choose preferences
for both musicality and prompt alignment. Each vote additionally includes a 1--5
confidence score. A rationale explaining the decision along the two dimensions is also included for future research.

CMI-Pref comprises 4,027 preference samples from generations produced by 23 models,
with a total duration of 133.8 hours. We reserve a balanced 500-pair test set,
with a 1:1:1:1 split across text, text+lyrics, text+audio, and text+audio+lyrics
conditions. Three annotators re-annotated all 500 test pairs; agreement with the
original labels reaches 75.2\% for musicality and 75.0\% for alignment. Detailed train/test
agreement is provided in \autoref{app:annotator-agreement}, and modality statistics
are provided in \autoref{app:data-statis}.
\autoref{tab:data-statis} highlights our distinct advantages in scale, duration,
and modality diversity over existing benchmarks.

\subsection{CMI-RewardBench}
We introduce CMI-RewardBench, a unified benchmark designed to evaluate the capability
of music reward models in capturing human aesthetic and instructional preferences. It uses only held-out data: \textbf{PAM} provides 500 scalar ratings
for musicality and text-music alignment; the test split of \textbf{MusicEval} provides 413
scalar musicality ratings; \textbf{Music Arena} provides 1,340 filtered pairwise
preferences; and the test split of \textbf{CMI-Pref} provides 500 pairwise musicality and prompt-alignment
labels under compositional conditions.

\subsubsection{Evaluation Protocol}

To ensure robust evaluation across heterogeneous label formats, we employ two
protocols: regression-based correlation and preference-based accuracy. For scalar
ratings in PAM and MusicEval, we report {Linear Correlation Coefficient (LCC)},
{Spearman Rank Correlation (SRCC)}, and {Kendall-Tau (K-Tau)}. For pairwise
labels in Music Arena and CMI-Pref, models must select the preferred audio, and
we report accuracy against human annotations.

\begin{figure*}[h]
  \vskip 0.2in
  \begin{center}
    \centerline{\includegraphics[width=1.3\columnwidth]{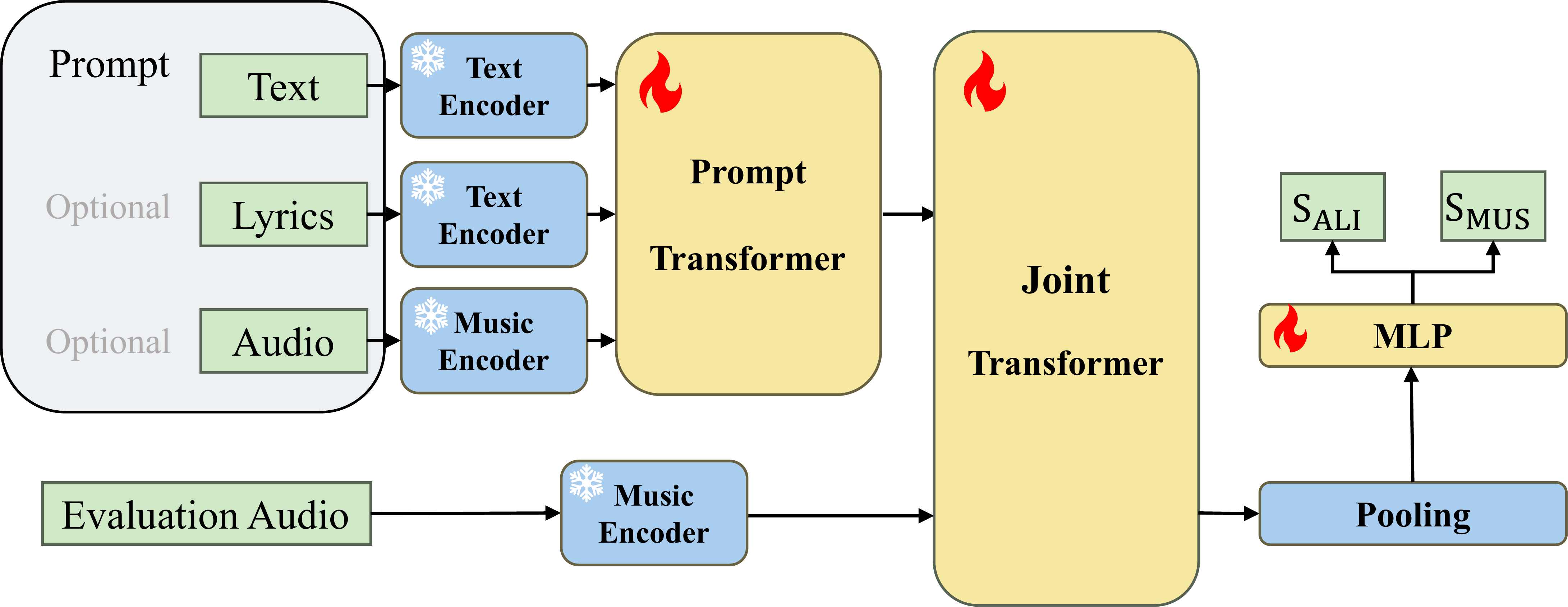}}
    \caption{ Model architecture of compositional music instruction reward model
      (CMI-RM).
    }
    \label{fig:model-architecture}
  \end{center}
\end{figure*}
\subsubsection{Baseline Models}
We benchmark a diverse set of current state-of-the-art models, categorized by their
primary training objective. The musicality baselines includes specialized MOS predictors such as PAM \cite{pam}, audiobox
\cite{audiobox} and SongEval \cite{songeval}. PAM is a reference-free metric
leveraging audio-language models for general-purpose quality assessment. Audiobox-Aesthetic
is an assessment framework providing four specialized metrics: Production
Quality (PQ), Production Complexity (PC), Content Enjoyment (CE), and Content Usefulness
(CU). SongEval is a model specifically trained on full-length songs using five
aesthetic dimensions: coherence, memorability, naturalness, clarity, and overall
musicality.

For text-music alignment, we examine similarity-based metrics: We use CLAP-Score in both default \citep{wu2023large}
and suggested music-specialized checkpoints \citep{grotschla2025benchmarking},
as well as CLAMP3 \citep{wu2025clamp3universalmusic}, a state-of-the-art multimodal
alignment model for ABC symbolic notation, waveform and text. MuQ-MuLan\citep{zhu2025muqselfsupervisedmusicrepresentation}
is a joint music-text embedding framework utilizing self-supervised music
representation learning.

We also include a general-purpose open-source reward model, Omni-Reward
\citep{jin2025omnirewardgeneralistomnimodalreward}.
Furthermore, we evaluate the zero-shot capabilities of frontier AudioLLMs for
all three tasks: musicality, text-music alignment, and compositional instruction
alignment. The evaluated suite includes the Qwen series and Gemini series: Qwen2-Audio
\citep{chu2024qwen2}, Qwen2.5-Omni \citep{xu2025qwen2}, and Qwen3-Omni, Gemini
2.5 Flash, 2.5 Pro \citep{comanici2025gemini}, and 3 Pro.
\subsection{Compositional Music Reward Modeling}

We develop a reward model architecture to handle multimodal conditioning and predict
fine-grained human scores.

\subsubsection{Model Architecture} %

\paragraph{Task Formulation.}
Given a compositional prompt
\[
  \mathcal{P}= (t, l, a_{\text{ref}}),
\]
where $t$ denotes optional text description, $l$ optional lyrics, and
$a_{\text{ref}}$ optional reference audio, together with an evaluation audio $a
    _{\text{eval}}$, our reward model predicts human-aligned preferences over
generated music along the two complementary dimensions {musicality (MUS)} and
  {alignment (ALI)}. The model outputs two scalar scores $(s_{\text{MUS}}, s_{\text{ALI}}
  ) \in \mathbb{R}^{2}$.

\paragraph{Architecture.}
We adopt a two-tower multimodal architecture following \citet{zhang2024hiveharnessinghumanfeedback}.
One tower encodes the multimodal prompt, while the other processes the target audio.
All encoders are frozen and instantiated from MuQ-MuLan. Text descriptions $t$
and lyrics $l$ are encoded separately using the text encoder, while reference
audio $a_{\text{ref}}$ and evaluation audio $a_{\text{eval}}$ are encoded
using the audio encoder. Empty modalities are treated as zero tensors during training
and inference. The encoded prompt components are concatenated and fused using
a 4-layer Prompt Transformer:
\begin{equation}
  \mathbf{h}_{\text{prompt}}= \text{PromptTF}([\mathbf{E}_{t}; \mathbf{E}_{l};
    \mathbf{E}_{a_{\text{ref}}}]),
\end{equation}
where $[\cdot;\cdot]$ denotes sequence concatenation.

To model interactions between the prompt and the generated music, the fused
prompt embedding and the evaluation audio embedding are concatenated and
processed by a single-layer self-attention Joint Transformer:
\begin{equation}
  \mathbf{h_{\text{prompt}}}; \mathbf{h}_{\text{eval}}= \text{JointTF}([\mathbf{h}
    _{\text{prompt}}; \mathbf{E}_{a_{\text{eval}}}]).
\end{equation}

We extract the hidden states corresponding to the evaluation audio tokens, apply
temporal pooling, and project them through a lightweight MLP to obtain the
final scores:
\begin{equation}
  (s_{\text{ALI}}, s_{\text{MUS}}) = \text{MLP}(\text{Pool}(\mathbf{h}_{\text{eval}}
  )).
\end{equation}

\subsubsection{Training Strategy}

We train the reward model using a two-stage pipeline that leverages both large-scale
pseudo-labeled data and high-quality human annotations. The datasets involved
include CMI-Pref-Pseudo for preference pre-training, together with the training
split of CMI-Pref and the training split of MusicEval for expert fine-tuning.
Both MUS and ALI heads are optimized jointly throughout both stages.
Given a training sample, we compute the musicality-related loss
$\mathcal{L}_{\text{MUS}}$ and the alignment-related loss
$\mathcal{L}_{\text{ALI}}$ when applicable. The overall training objective is an
average of the two:
\begin{equation}
  \mathcal{L}_{\text{total}}= \frac{1}{2}( \mathcal{L}_{\text{MUS}}+\mathcal{L}
  _{\text{ALI}})
\end{equation}

\paragraph{Stage 1: Preference Pre-training.}
We first pre-train the model on preference pairs from CMI-Pref-Pseudo. Training
is conducted for 2k steps with a batch size of 48.
Pairwise preferences are modeled using Bradley--Terry \cite{bradley1952rank}
formulation. Given a prompt $\mathcal{P}$ and two candidate audio files $A$ and
$B$,
\begin{equation}
  P(A > B) = \sigma\big(s_{\theta}(\mathcal{P}, A) - s_{\theta}(\mathcal{P}, B)
  \big),
\end{equation}
where $s_{\theta}$ denotes the predicted MUS or ALI score depending on the
annotation type. The model is optimized using cross-entropy loss, and tied
preferences are excluded from training. To mitigate over-confident decision boundaries
induced by noisy pseudo labels, we apply label smoothing with a ratio of 0.2
during this stage. Furthur explanation can be found in Appendix \ref{Append:Label
  Smoothing}.

\paragraph{Stage 2: Expert Fine-tuning.}
We then fine-tune the model on a mixture of high-quality human annotations, combining
the training split of {CMI-Pref} and {MusicEval}, resulting in a total of 6{,}647
training samples. We use a batch size of 48 and perform early stopping based on
validation performance. The selected checkpoint is obtained after 250
optimization steps selected with early stopping.
Human annotations appear in two formats: (1) pairwise preferences
$(\mathcal{P}, A, B)$, trained using Bradley--Terry loss as in Stage~1; and (2)
scalar ratings $(\mathcal{P}, A, y)$, where $y \in [1, 5]$. For scalar ratings,
we regress the predicted scores using
\begin{equation}
  \mathcal{L}_{\text{reg}}= \text{MSE}\big(2 \tanh(a s + b) + 3,\; y\big),
\end{equation}
where $s$ denotes the predicted MUS or ALI score. The scaling parameters are initialized
to $a=0.2$ and $b=0$ during fine-tuning. We drop these constants during
inference.

\subsubsection{Test-time Scaling}

To evaluate the efficacy of test-time scaling \cite{jin2025inferencetime}, we
conduct experiments using two backbone models: \textbf{MusicGen-small} and
\textbf{Stable-Audio-Open-small}. For each of the 2,183 text prompts from the
MusicCaps \cite{agostinelli2023musiclm} dataset eval-split, we generate 10 audio
samples (10 sec each) per model. Our reward model serves as a ``best-of-N''
filter to select the top-performing sample, where $N \in \{1, 3, 10\}$. We
evaluate the effectiveness via subjective A/B testing: Validating whether reward-model
selection consistently aligns with human preferences for superior musical quality.
The selection criterion is the average of musicality and alignment scores. Details of
the subjective test can be found in Appendix~\ref{app:stat_testtime}.

\section{Discussion}

\subsection{Benchmark Results}
\subsubsection{Comprehensive Evaluation on Musicality}

\begin{table*}
  [h]
  \caption{Musicality results on CMI-RewardBench. For each metric, the best
    performance is marked in boldface and second with underline.}
  \label{tab:musicality-all-data-statis}
  \centering
  \scriptsize \resizebox{\textwidth}{!}{%
    \begin{tabular}{l|ccc|ccc|cc}
      \toprule \multirow{1}{*}{Musicality}     & \multicolumn{3}{c|}{PAM (Music Subset)} & \multicolumn{3}{c|}{MusicEval (Test Split)} & \makecell[c]{Music Arena } & CMI-Pref                                                                                                 \\
      Method\&Model                            & LCC                                     & SRCC                                        & K-Tau                      & LCC                & SRCC               & K-Tau              & ACC                 & ACC                 \\
      \midrule PAM score                       & 0.5873                                  & 0.6099                                      & 0.4367                     & 0.6466             & 0.6724             & 0.4874             & 63.13\%             & 65.40\%             \\
      audiobox-CE                              & 0.5283                                  & 0.5204                                      & 0.3665                     & 0.6393             & 0.6599             & 0.4830             & 64.25\%             & 71.80\%             \\
      audiobox-CU                              & 0.4645                                  & 0.4704                                      & 0.3279                     & 0.6272             & 0.6764             & 0.4950             & 67.76\%             & 71.40\%             \\
      audiobox-PC                              & 0.2505                                  & 0.2230                                      & 0.1552                     & 0.1225             & 0.0768             & 0.0514             & 58.73\%             & 59.00\%             \\
      audiobox-PQ                              & 0.4636                                  & 0.4513                                      & 0.3166                     & 0.6016             & 0.6335             & 0.4620             & 67.54\%             & 73.80\%             \\
      SongEval-RM                              & \textbf{0.6987}                         & \underline{0.6977}                          & \underline{0.4997}         & 0.7140             & 0.6949             & 0.5185             & \textbf{73.88\%}    & 72.40\%             \\
      Omni-Reward                              & 0.3364                                  & 0.3115                                      & 0.2128                     & 0.5306             & 0.5137             & 0.3642             & 54.03\%             & 65.60\%             \\
      \midrule Qwen2-audio                     & 0.1468                                  & 0.1523                                      & 0.1120                     & 0.1455             & 0.2196             & 0.1585             & 5.99\%              & 8.60\%              \\
      Qwen2.5-omni                             & 0.2776                                  & 0.2837                                      & 0.2144                     & 0.1655             & 0.1454             & 0.1145             & 36.05\%             & 17.40\%             \\
      Qwen3-omni                               & 0.4155                                  & 0.4113                                      & 0.3146                     & 0.3693             & 0.3101             & 0.2205             & 59.63\%             & 60.40\%             \\
      Gemini2.5-flash                          & 0.3813                                  & 0.3693                                      & 0.2571                     & 0.4188             & 0.3886             & 0.2694             & 64.12\%             & 64.20\%             \\
      Gemini2.5-pro                            & 0.4463                                  & 0.4355                                      & 0.3068                     & 0.4966             & 0.4902             & 0.3454             & 69.75\%             & 70.00\%             \\
      Gemini3-pro                              & 0.5972                                  & 0.5967                                      & 0.4283                     & 0.6044             & 0.6018             & 0.4400             & 68.85\%             & 65.80\%             \\
      \midrule - w/o f.t.: Distill only        & 0.4358                                  & 0.4304                                      & 0.2981                     & 0.5253             & 0.5117             & 0.3711             & 64.25\%             & 70.80\%             \\
      \makecell[l]{- w/ f.t.: CMI-Pref}        & \underline{0.6932}                      & \textbf{0.6988}                             & \textbf{0.5101}            & \underline{0.7272} & \underline{0.7315} & \underline{0.5495} & 71.41\%             & \underline{77.80\%} \\
      \makecell[l]{- w/ f.t.: CMI + MusicEval} & 0.6367                                  & 0.6606                                      & 0.4754                     & \textbf{0.8195}    & \textbf{0.8266}    & \textbf{0.6459}    & \underline{73.43\%} & \textbf{ 78.20\%}   \\
      \bottomrule
    \end{tabular}%
  } \vskip -0.1in
\end{table*}

\textbf{Superior Generalization on Musicality.} \autoref{tab:musicality-all-data-statis}
presents a quantitative comparison of our proposed \textbf{CMI-RM} against all
baselines. Our method demonstrates robust generalization capabilities across diverse
regression tasks. Specifically, the model fine-tuned on CMI-Pref achieves state-of-the-art
performance on the PAM music subset (e.g., SRCC of \textbf{0.6988}). Notably,
while specialized baselines such as SongEval-RM exhibit strong performance on
the Music Arena platform, their efficacy degrades on the more compositionally complex
CMI-Pref dataset (72.40\%). In contrast, by incorporating external data (w/ f.t.:
CMI + MusicEval), our CMI-RM maintains highly competitive accuracy on Music
Arena (\underline{73.43\%}) while achieving state-of-the-art preference alignment
on CMI-Pref (\textbf{78.20\%}). This indicates that our preference alignment strategy
yields robust representations that consistently correlate with granular human ratings
across varying evaluation scenarios.

\textbf{Deficiency of General-Purpose MLLMs.} Frontier MLLMs remain weaker for
fine-grained music judgment: on CMI-Pref musicality, Gemini 3 Pro and
Qwen3-omni obtain 65.80\% and 60.40\%, versus \textbf{78.20\%} for CMI-RM.
Some AudioLLMs (e.g., Qwen2-audio) often fail the required decision format,
leading to near-random results.

\subsubsection{Compositional MultiModal Instruction Alignment Evaluation}

\begin{table*}
  [ht]
  \caption{Benchmark Results on Compositional Multimodal Instruction Alignment.}
  \label{tab:text-music-alignment-single}
  \centering
  \scriptsize
  \setlength{\tabcolsep}{2pt}
  \renewcommand{\arraystretch}{1.15}
  \resizebox{0.9\textwidth}{!}{%
    \begin{tabular}{l|ccc|cccc|c|c}
      \toprule                                 & \multicolumn{3}{c|}{PAM (Music Subset)} & \multicolumn{6}{c}{CMI-Pref (Subsets)}                                                                                                                                                                            \\
      \midrule Text-Music                      & \CHECK                                  & \CHECK                                 & \CHECK             & \CHECK              & \CHECK              & \CHECK              & \CHECK              & \multirow{3}{*}{\makecell[l]{CMI-Pref                       \\ w/o Audio}} & \multirow{3}{*}{\makecell[l]{CMI-Pref \\ w/ Audio}} \\
      Lyrics-Music                             & \CROSS                                  & \CROSS                                 & \CROSS             & \CROSS              & \CHECK              & \CROSS              & \CHECK              &                                                             \\
      Audio-Music                              & \CROSS                                  & \CROSS                                 & \CROSS             & \CROSS              & \CROSS              & \CHECK              & \CHECK              &                                                             \\
      \midrule Metrics                         & LCC                                     & SRCC                                   & K-Tau              & ACC                 & ACC                 & ACC                 & ACC                 & ACC                                   & ACC                 \\
      \midrule CLAP score (default)            & 0.4692                                  & 0.4517                                 & 0.3171             & 60.80\%             & 64.00\%             & -                   & -                   & 62.40\%                               & -                   \\
      CLAP score (music)                       & 0.3192                                  & 0.2881                                 & 0.1978             & 67.20\%             & \underline{73.60}\% & -                   & -                   & \textbf{70.40\%}                      & -                   \\
      MuQ-Mulan                                & 0.4984                                  & 0.4741                                 & 0.3341             & 64.80\%             & 68.00\%             & -                   & -                   & 66.40\%                               & -                   \\
      CLAMP3 score                             & 0.2998                                  & 0.3013                                 & 0.2068             & 63.20\%             & 62.40\%             & -                   & -                   & 62.80\%                               & -                   \\
      Omni-Reward                              & 0.3376                                  & 0.3072                                 & 0.2120             & 56.80\%             & \textbf{76.80\%}    & 67.20\%             & 70.40\%             & 66.80\%                               & 68.80\%             \\
      \midrule Qwen2-audio                     & -0.024                                  & -0.025                                 & -0.020             & 0.80\%              & 2.40\%              & 8.80\%              & 14.40\%             & 1.60\%                                & 11.60\%             \\
      Qwen2.5-Omni                             & 0.1529                                  & 0.2084                                 & 0.1696             & 31.20\%             & 37.60\%             & 25.60\%             & 32.00\%             & 34.40\%                               & 28.80\%             \\
      Qwen3-Omni                               & \textbf{0.5841}                         & \textbf{0.5907}                        & \textbf{0.4714}    & 67.20\%             & 60.00\%             & 64.80\%             & 63.20\%             & 63.60\%                               & 64.00\%             \\
      Gemini2.5-Flash                          & 0.3686                                  & 0.2454                                 & 0.1851             & 65.60\%             & 56.00\%             & 69.60\%             & 54.40\%             & 60.80\%                               & 62.00\%             \\
      Gemini2.5-Pro                            & 0.4562                                  & 0.4179                                 & 0.3192             & \textbf{71.20\%}    & 63.20\%             & 73.60\%             & 72.00\%             & 67.20\%                               & 72.80\%             \\
      Gemini3-Pro                              & \underline{0.5201}                      & \underline{0.5373}                     & \underline{0.4047} & 67.20\%             & 60.80\%             & 68.80\%             & 64.80\%             & 64.00\%                               & 66.80\%             \\
      \midrule - w/o f.t.: Distill only        & 0.3647                                  & 0.3547                                 & 0.2449             & 67.20\%             & 72.00\%             & 69.20\%             & 77.20\%             & 69.60\%                               & 73.20\%             \\
      \makecell[l]{- w/ f.t.: CMI-Pref}        & 0.5200                                  & 0.5243                                 & 0.3721             & 64.80\%             & 72.80\%             & \underline{76.00}\% & \textbf{82.40}\%    & 68.80\%                               & \textbf{79.20\%}    \\
      \makecell[l]{- w/ f.t.: CMI + MusicEval} & 0.4418                                  & 0.4321                                 & 0.3008             & \underline{67.60}\% & 72.80\%             & \textbf{76.40}\%    & \underline{79.20}\% & \underline{70.20\%}                   & \underline{77.80\%} \\
      \bottomrule
    \end{tabular}%
  } \vskip -0.1in
\end{table*}

\autoref{tab:text-music-alignment-single} displays the compositional alignment
evaluation. Unlike standard benchmarks that primarily focus on text-to-music
consistency, CMI-RewardBench utilizes the CMI-Pref dataset to evaluate the capability
of models to follow complex instructions involving text, lyrics, and reference
audio simultaneously.

\textbf{Comparison with Objective Metrics and General MLLMs.} Standard metrics
and general MLLMs are limited under compositional conditions. CLAP (default)
generally yield moderate performance (about 60--64\% on CMI-Pref subsets) but cannot handle
composed audio-conditioned cases. Qwen3-Omni is strong on PAM (LCC 0.5841) but weaker on
CMI-Pref w/ audio (64.0\%). CMI-RM reaches \textbf{70.20\%} on CMI-Pref w/o
audio (CMI+MusicEval) and \textbf{79.20\%} on w/ audio (CMI-Pref fine-tuned),
surpassing Gemini 2.5 Pro (67.2\% and 72.8\%). This validates that CMI-RM provides
a unified solution for diverse and multimodal generation conditions.

\textbf{Detailed Breakdown by Modality.} On the hardest Text+Lyrics+Audio
subset, CMI-Pref fine-tuning reaches \textbf{82.40\%}, well above Gemini 3 Pro
(66.8\%). For w/o-audio subsets, CMI+MusicEval gives the best result
(\textbf{70.20\%}), indicating complementary gains from MusicEval.

\subsection{The Effectiveness of Different Training Sets}
\label{sec:training_set_ablation}

We ablate how training data sources affect CMI-RM under a fixed architecture
and identical trainable parameters. All variants share the same two-head setup
(MUS/ALI) and differ only in (i) initialization and (ii) fine-tuning data.

\paragraph{Training variants.}
\begin{itemize}[leftmargin=1.2em, itemsep=0.1em, topsep=0.1em, parsep=0pt]
  \item \textbf{Distill only}: Pre-trained on CMI-Pref-Pseudo only (pairwise
        preference, 1.5 epochs).
  \item \textbf{Distill+CMI-Pref}: Initialized from \textbf{Distill}, then fine-tuned
        on CMI-Pref (train split).
  \item \textbf{Distill+MusicEval}: Initialized from \textbf{Distill}, then fine-tuned
        on MusicEval (train split).
  \item \textbf{Distill+Both}: Initialized from \textbf{Distill}, then jointly
        fine-tuned on CMI-Pref + MusicEval.
  \item \textbf{Scratch+Both}: Random initialization, trained on CMI-Pref +
        MusicEval.
\end{itemize}

\paragraph{Aggregated metric.}
To summarize results across heterogeneous benchmarks, we report one aggregated
score per dataset in Table~\ref{tab:ablation}. For PAM, we average SRCC over the
\emph{musicality} and \emph{text--music alignment} regression tasks. For
MusicEval, we report SRCC on its musicality MOS. For Music Arena, we report the
accuracy of the preference, where the prediction of the model uses the
difference between the musicality score $s_{\text{MUS}}$ of the pair. For CMI-Pref,
we report mean accuracy averaged over musicality and alignment preferences. Detailed
per-task numbers are provided in Table~\ref{tab:musicality-all-data-statis},
Table~\ref{tab:text-music-alignment-single}, and Table~\ref{tab:ablation}.

\begin{table}[H]
  \small
  \centering
  \caption{ Aggregated ablation results with a fixed CMI-RM architecture. Mean
    SRCC on PAM averages the \emph{musicality} and \emph{text--music alignment}
    regression tasks. Music Arena accuracy uses musicality score $s_{\text{MUS}}$
    for pairwise prediction. Mean Acc.\ on CMI-Pref averages musicality and alignment
    preference accuracy. }
  \label{tab:ablation} \resizebox{\columnwidth}{!}{
    \begin{tabular}{lcccc}
      \toprule \textbf{Training variant} & \textbf{PAM}       & \textbf{MusicEval} & \textbf{Music Arena} & \textbf{CMI-Pref}   \\
      \textbf{Metric} $\uparrow$         & Mean SRCC          & SRCC               & Acc.                 & Mean Acc.           \\
      \midrule Distill only              & 0.3925             & 0.5117             & 64.25\%              & 71.10\%             \\
      Distill+CMI-Pref                   & \textbf{0.6116}    & \underline{0.7315} & \underline{71.41\%}  & \underline{75.90\%} \\
      Distill+MusicEval                  & 0.4338             & 0.3460             & 64.78\%              & 69.00\%             \\
      Distill+Both                       & \underline{0.5464} & \textbf{0.8266}    & \textbf{73.43\%}     & \textbf{76.05\%}    \\
      Scratch+Both                       & 0.2630             & 0.4986             & 71.34\%              & 72.15\%             \\
      \bottomrule
    \end{tabular}
  }
\end{table}

\paragraph{Findings.}

\textbf{(1) CMI-Pref is the primary driver for cross-benchmark generalization.}
Fine-tuning on CMI-Pref consistently improves all four aggregated metrics: \textbf{Distill+CMI}
yields consistent improvements over the \textbf{Distill} baseline on PAM (0.6116
vs.\ 0.3925), MusicEval (0.7315 vs.\ 0.5117), Music Arena (71.41\% vs.\ 64.25\%),
and CMI-Pref (75.90\% vs.\ 71.10\%). This suggests that high-quality human preference
data, when coupled with compositional conditions, provides highly transferable
supervision signals.

\textbf{(2) MusicEval provides a complementary signal, but is insufficient alone.}
While \textbf{Distill+MusicEval} alone does not yield consistent gains across all
sets, \textbf{Distill+Both} effectively integrates both datasets. This joint
training substantially improves MusicEval correlation (to \textbf{0.8266}) while
pushing Music Arena and CMI-Pref to their peak performances (\textbf{73.43\%} and
\textbf{76.05\%}, respectively). We observe a mild trade-off on PAM compared to
\textbf{Distill+CMI}, indicating a slight objective mismatch between PAM and MusicEval
that the joint training mechanism partially mitigates, while more diverse data
is required for better generalization.

\textbf{(3) Distillation initialization is universally beneficial.} Comparing \textbf{Distill+Both}
with \textbf{Scratch+Both}, we observe that distillation initialization yields
consistent gains across all benchmarks. Most notably, it improves performance
on Music Arena from 71.34\% to \textbf{73.43\%}, and significantly boosts both
PAM (0.2630 to 0.5464) and CMI-Pref (72.15\% to \textbf{76.05\%}). These
results suggest that our large-scale pseudo-label pre-training (\textbf{Distill})
establishes a robust prior that not only mitigates overfitting but also transfers
effectively to noisy, out-of-distribution scenarios such as Music Arena, thereby
establishing \textbf{Distill+Both} as the optimal training strategy.

\textbf{(4) Independence on Qwen3-Omni pseudo labels.}
Our RM is not simply copying the pseudo-labeler. On 500 CMI-Pref-test samples,
The agreement rate for two inferences of Qwen-Omni3 is 94.4\% musicality / 94.1\% alignment,
but its agreement with our pseudo-pretrained model after label smoothing  is only 80.6\% / 79.2\%, and drops to 63.7\% / 68.4\% after human finetuning.
This shows human supervision shifts the decision boundary away from the teacher.
Distill+CMI-Pref and Distill+Both also outperform Qwen3-Omni on MusicArena, improving accuracy from 59.63\% to 71.41\% with 1,340 human labels.
Adding a second pseudo source (40k Gemini samples) drops the model agreement of the pretrained checkpoint with Qwen3-Omni to 65.2\% and 67.3\%, but yields only minor final gains shown in \autoref{tab:mixed_pseudo_ablation}.

\subsection{Test-Time Scaling}

\begin{figure}[t]
  \centering
  \setlength{\tabcolsep}{0pt}

  \begin{subfigure}
    {\columnwidth}
    \centering
    \includegraphics[width=0.8\columnwidth]{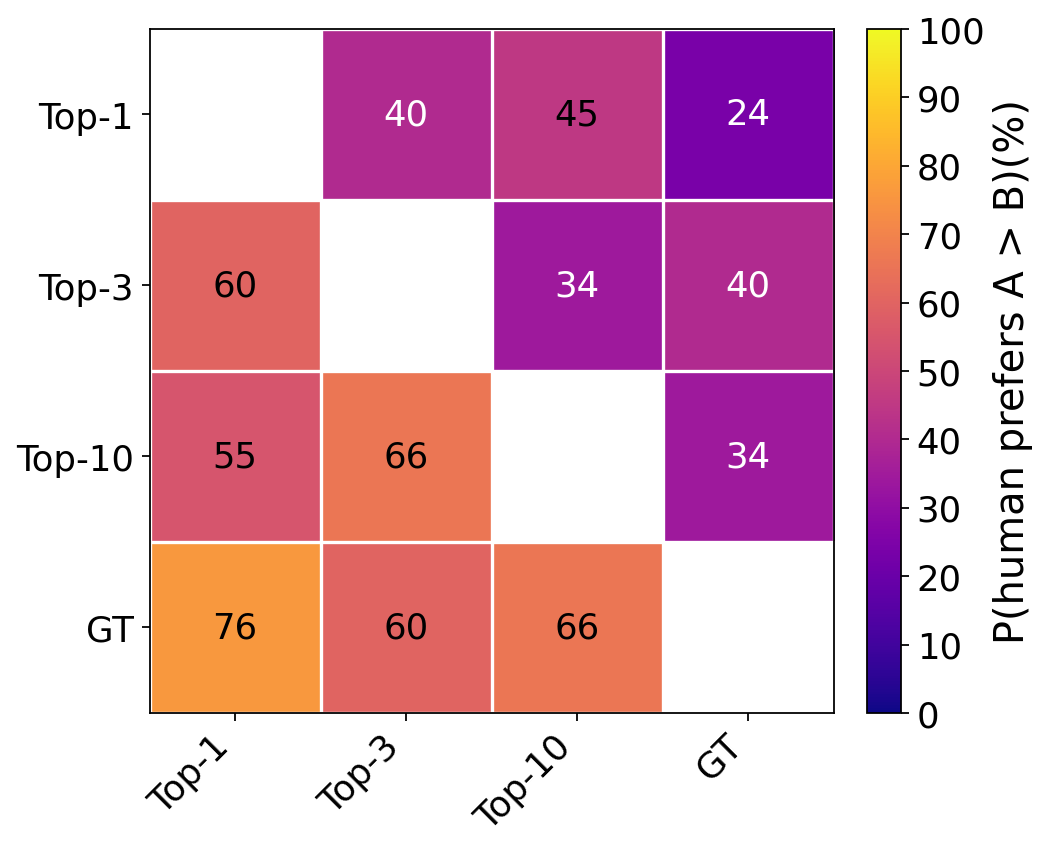}
    \caption{MusicGen-small}
    \label{fig:heatmap_musicgen}
  \end{subfigure}

  \vspace{2mm}

  \begin{subfigure}
    {\columnwidth}
    \centering
    \includegraphics[width=0.8\columnwidth]{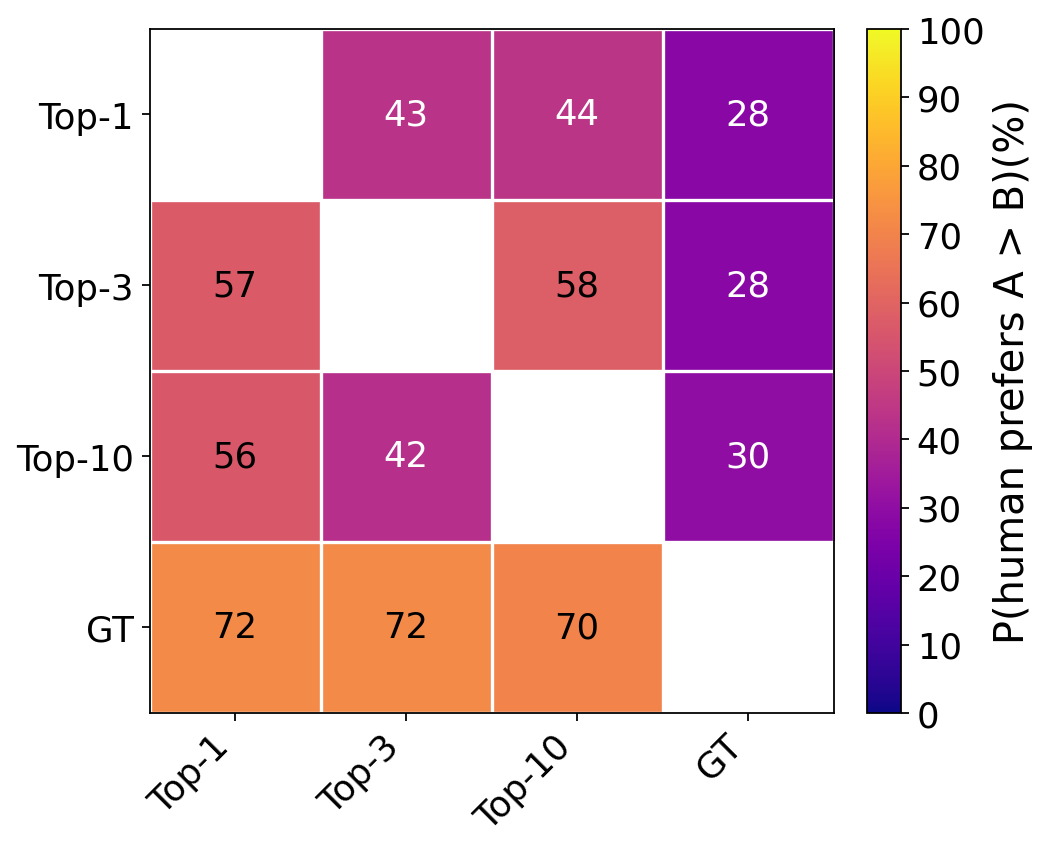}
    \caption{Stable-Audio-Open-small}
    \label{fig:heatmap_stableaudio}
  \end{subfigure}

  \caption{ Pairwise preference matrices for test-time scaling with RM reranking.
    Each cell reports the percentage of trials in which annotators preferred
    system $A$ (row) over system $B$ (column). }
  \label{fig:heatmaps}
\end{figure}

Table~\ref{tab:testtime-all} shows that RM-based best-of-$N$ reranking provides
consistent test-time scaling gains across both backbones. The improvements are
more pronounced for \textbf{MusicGen-small}, where MUQ-MULAN and AudioBox/SongEval
metrics increase monotonically from $N{=}1$ to $N{=}10$. For \textbf{Stable-Audio-Open-small},
the gains are smaller and begin to saturate, suggesting diminishing returns
when candidate quality is concentrated.

The human preference matrices in Fig.~\ref{fig:heatmaps} provide two takeaways.
First, \textbf{Ground Truth (GT) is consistently preferred over all reranked
  outputs} for both backbones, indicating that reranking improves generation quality
but does not close the gap to real data. Second, preferences among Top-$k$ selections
are \emph{not strictly monotonic} with $N$, and the gains can saturate (e.g.,
Stable Audio shows limited differences between $N{=}3$ and $N{=}10$). Overall,
best-of-$N$ reranking is a simple and effective test-time scaling strategy,
with diminishing returns beyond moderate $N$.

\begin{table}[t]
  \caption{Test-time scaling results of MusicGen-small and Stable-audio-open-small
    on objective metrics on text-music alignment (MuQ-Mulan) and musicality (AudioBox,
    and SongEval).}
  \label{tab:testtime-all} \vskip 0.05in
  \centering
  \small
  \resizebox{\columnwidth}{!}{%
    \begin{tabular}{lc|cc|c}
      \toprule \multirow{2}{*}{Model} & \multicolumn{1}{c|}{Alignment} & \multicolumn{2}{c|}{AudioBox} & \multicolumn{1}{c}{SongEval}                 \\
                                      & MuQ-MuLan$\uparrow$            & CE$\uparrow$                  & CU$\uparrow$                 & Mus$\uparrow$ \\
      \midrule MusicGen               & 0.298                          & 6.046                         & 6.989                        & 2.143         \\
      MusicGen (N=3)                  & 0.323                          & 6.405                         & 7.255                        & 2.213         \\
      MusicGen (N=10)                 & 0.339                          & 6.647                         & 7.416                        & 2.273         \\
      \midrule Stable Audio           & 0.293                          & 5.567                         & 7.170                        & 2.055         \\
      Stable Audio (N=3)              & 0.301                          & 5.732                         & 7.245                        & 2.078         \\
      Stable Audio (N=10)             & 0.307                          & 5.799                         & 7.290                        & 2.090         \\
      \bottomrule
    \end{tabular}
  }
  \vskip -0.08in
\end{table}

\section{Conclusion}

We introduce \textbf{CMI-RewardBench}, a unified benchmark for reward modeling
under compositional multimodal instruction (text, lyrics, and reference audio),
together with two datasets: \textbf{CMI-Pref-Pseudo} (110k pseudo-labeled
pairs) and \textbf{CMI-Pref} (4k expert annotations with confidence).

By integrating CMI-Pref with existing resources (e.g., PAM, MusicEval, Music Arena),
CMI-RewardBench provides a rigorous testbed spanning five tasks from absolute scoring
to pairwise preference, revealing a clear capability gap that even frontier
multimodal LLM LLM judges struggle to reach strong agreement with expert
preferences in this setting.

We further developed \textbf{CMI-RM}, a parameter-efficient reward model that
supports compositional conditioning over text, lyrics, and audio within a
single architecture, achieving performance competitive with or exceeding specialized
open-source baselines, and providing measurable gains when used for best-of-$N$
reranking as a simple inference-time scaling strategy. We release the dataset,
benchmark, and model weights to facilitate future research.

\section*{Conflict of Interest Disclosure}
Yinghao Ma also acknowledges support from the Google PhD Fellowship.
Because this work includes an evaluation of the Gemini series, a family of models developed by Google,
this support is disclosed as a potential financial conflict of interest. The
authors declare no other financial conflicts of interest.

\section*{Acknowledgement}
Yinghao Ma is a research student at the UKRI Centre for Doctoral Training in Artificial
Intelligence and Music, supported by UK Research and Innovation [grant number EP/S022694/1].
Yinghao Ma also acknowledges the support of Google PhD Fellowship. 
Hewei Gao is a PhD student partially supported by alignAI, the alignAI Doctoral Network, funded by the European Union’s Marie Skłodowska-Curie Actions programme [grant number 101169473].
We thank Termeh Taheri on suggestions on back-end development and Christos Plachouras on feedbacks on platform design.

\section*{Impact Statement}

This work advances the evaluation and alignment of music generation systems by
providing a unified benchmark and lightweight reward models that better reflect
human preference under compositional multimodal conditions. Our pipeline
includes audio generated via commercial APIs; therefore we follow a TOS-aware release
policy: only components that are explicitly redistributable are made public, and
any restricted parts are shared (if at all) via an application-based access process.
We also acknowledge potential copyright concerns from style/melody similarity
in generated music and provide a takedown/correction mechanism. Human preference
data are collected via a Music-Arena-style platform under informed consent and
data minimization (no personal identifiers; limited metadata), with an opt-out/withdrawal
mechanism. Given previous datasets included in benchmark are all CC-BY or CC-NC
license, our datasets and benchmark are released under \textbf{CC-BY-NC-SA}
license and accompanied by a datasheet/data card documenting sources, licenses/TOS
compatibility, and the public release fields.

\newpage
\bibliography{main}
\bibliographystyle{icml2026}

\clearpage
\appendix

\section{Detailed Metadata of Datasets}
\label{app:data-statis}
\subsection{Diversity of Prompts}

On the basis of a text prompt, the retained CMI-Pref-Pseudo split covers audio+lyrics (18.3\%), audio-only
(17.0\%), lyrics-only (19.8\%), and text-only (44.8\%) conditions. CMI-Pref
contains audio+lyrics (15.2\%), audio-only (20.4\%), lyrics-only (14.9\%), and
text-only (49.5\%) conditions, with the 500-pair test split balanced across the
four prompt modalities (125 pairs each).

CMI-Pref contains a highly diverse collection of prompts and lyrics that reflect
realistic and heterogeneous user instructions for music generation. Across the
full dataset, we collect 10,213 unique prompts, while the human-annotated
split contains 2,788 unique prompts. Prompt lengths follow a long-tailed distribution:
most prompts are concise style or intent descriptors consisting of a few words,
while a non-trivial subset includes longer compositional instructions specifying
multiple musical attributes. This range captures both minimal user inputs and
more detailed requests that require structured reasoning from reward models.

Semantically, the prompts span a wide variety of musical attributes. They
cover a broad spectrum of genres, including popular, electronic, rock, jazz, classical,
ambient, folk, and orchestral styles, with many prompts combining multiple genre
cues. Beyond genre, prompts frequently specify mood, tempo, instrumentation,
and production characteristics, introducing orthogonal dimensions of variation.
Such compositional prompts prevent reward models from relying on single-keyword
correlations and instead encourage holistic assessment of instruction
adherence.

The prompt distribution is also linguistically diverse. While English prompts dominate,
the dataset includes non-English and mixed-language prompts, reducing reliance
on a single linguistic prior and better reflecting global usage patterns.
Importantly, prompts are paired with generations produced by a wide range of music
generation models and commercial APIs, ensuring that instruction diversity is
not confounded with a specific synthesis pipeline.

In addition to text prompts, CMI-Pref includes lyric-conditioned instructions at
a meaningful scale. The human-annotated split contains 840 examples with non-empty
lyrics, and the full dataset contains 3,896 such examples. Lyrics vary
substantially in structure and length, ranging from short repetitive hooks to multi-stanza
verses with clear narrative progression. They include both vocal-focused
instructions and lyrics intended to be adapted to different musical styles, posing
additional alignment challenges beyond text-to-music generation.

Overall, the diversity of prompts and lyrics in CMI-Pref provides a realistic
and challenging testbed for reward modeling. By combining short and long instructions,
multiple semantic control dimensions, diverse musical styles, and lyric-conditioned
inputs, the dataset supports robust evaluation of reward models under compositional
and multimodal instruction settings.

\subsection{Audio Duration and Listening-Time Statistics}
\label{app:duration-listening-stats} To better characterize both the temporal profile
of the generated samples and annotator behavior, we analyze three
complementary aspects: (i) the distribution of audio duration, (ii) the
distribution of total listening time, and (iii) the correlation between these two
quantities.

\subsubsection{Distribution of Audio Duration}
\begin{figure}[htbp]
  \centering
  \includegraphics[width=.96\columnwidth]{
    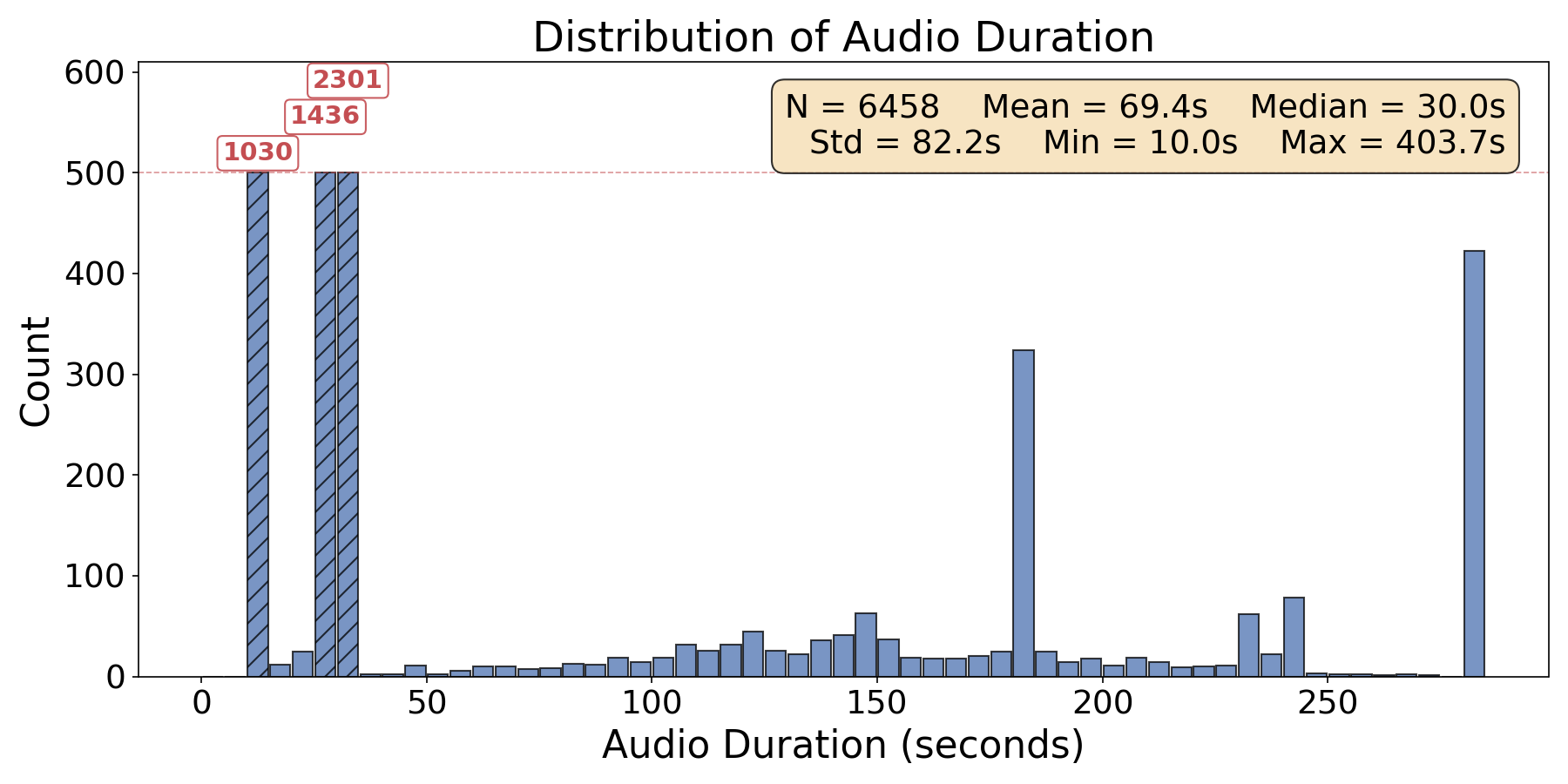
  }
  \caption{Distribution of audio duration. The x-axis denotes generated audio duration
    (seconds), and the y-axis denotes sample count ($N=6458$ unique audios). To improve
    readability, bar heights are clipped at 500 and overflow counts are
    annotated in red; a tiny fraction of samples with duration $>$300 seconds is
    omitted from the plot.}
  \label{fig:audio_duration_distribution}
\end{figure}

Figure~\ref{fig:audio_duration_distribution} reports the duration distribution
of unique generated audios ($N=6458$). The distribution is multimodal: while the
dominant mass is concentrated in the short-duration region (especially around
30 seconds), additional peaks appear at longer fixed durations. This pattern is
consistent with heterogeneous generation constraints across source models,
where many earlier systems primarily support short outputs while others use
longer default lengths. Overall, the mean ($69.4$ seconds) remains notably
higher than the median ($30.0$ seconds), indicating a long right tail.

For readability, we cap the displayed histogram height at 500 counts and
annotate overflow bins in red (e.g., 1030, 1436, and 2301). In addition, a
very small number of samples with duration above 300 seconds are omitted from
visualization; the plotted subset still covers more than 99.5\% of all samples.

\subsubsection{Distribution of Total Listening Time}
We define \emph{total listening time} as the accumulated playback time an annotator
spends on one audio within a comparison task. Since CMI-Pref contains 4,027
pairwise votes and each vote includes two audios, we obtain 8,054 audio-level listening-time
records. We exclude 8 extreme records with listening time above 500 seconds (likely
idle-page artifacts), resulting in 8,046 valid samples for analysis.

\begin{figure}[htbp]
  \centering
  \includegraphics[width=.96\columnwidth]{
    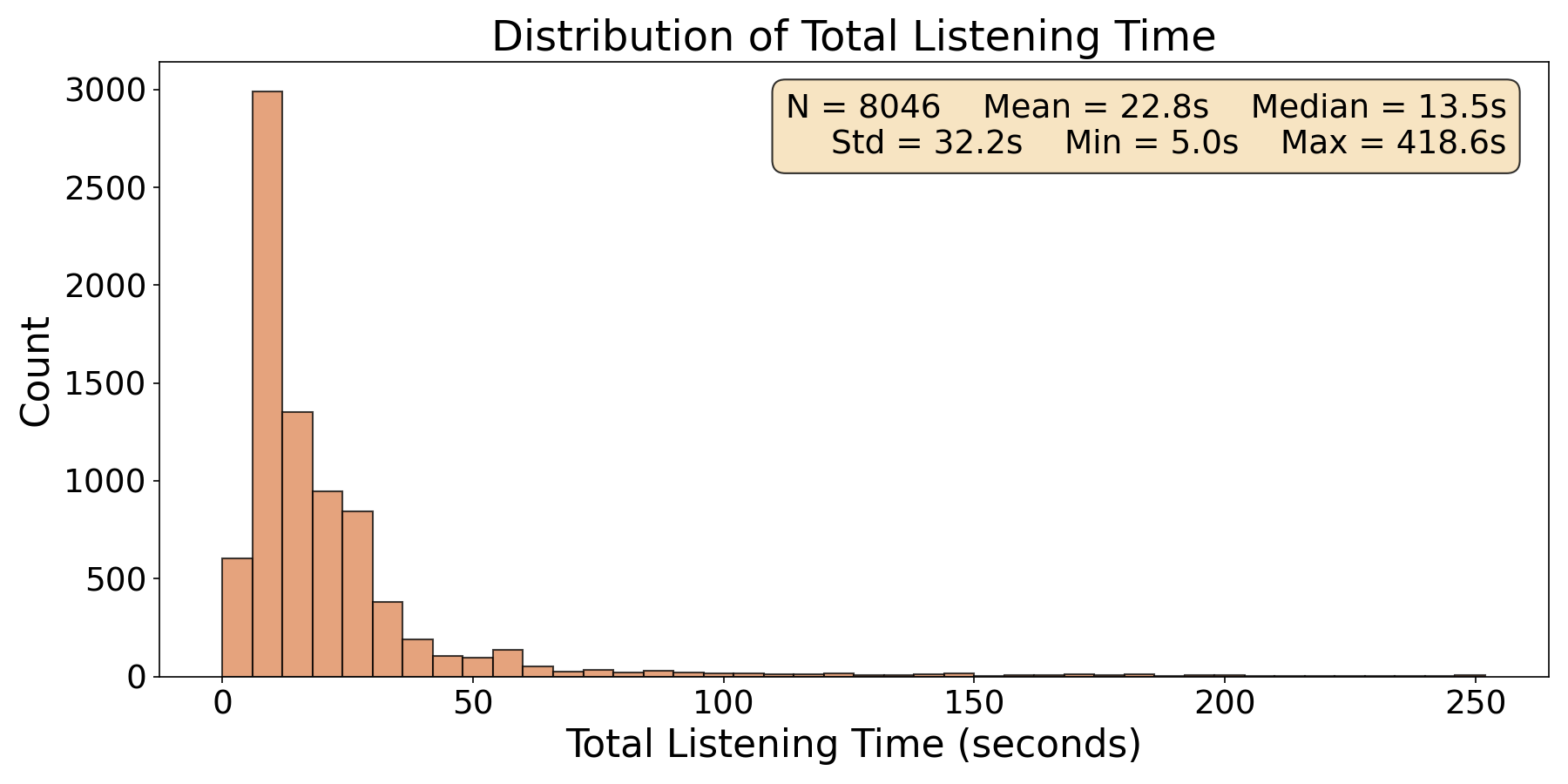
  }
  \caption{Distribution of total listening time per audio. ``Total listening time''
    denotes accumulated playback time spent by annotators on one audio. Starting
    from 8,054 records (4,027 votes $\times$ 2 audios), we remove 8 outliers with
    listening time $>$500 seconds and report the remaining $N=8046$ samples.}
  \label{fig:total_listening_time_distribution}
\end{figure}

As shown in Figure~\ref{fig:total_listening_time_distribution}, listening time
is also long-tailed, with mean $22.8$ seconds and median $13.5$ seconds. The
majority of samples fall within 50 seconds, and most are above 10 seconds.
This pattern suggests that annotators typically listen long enough to form
meaningful pairwise judgments, while keeping the annotation process efficient.

\subsubsection{Correlation Between Audio Duration and Total Listening Time}
\begin{figure}[htbp]
  \centering
  \includegraphics[width=.96\columnwidth]{
    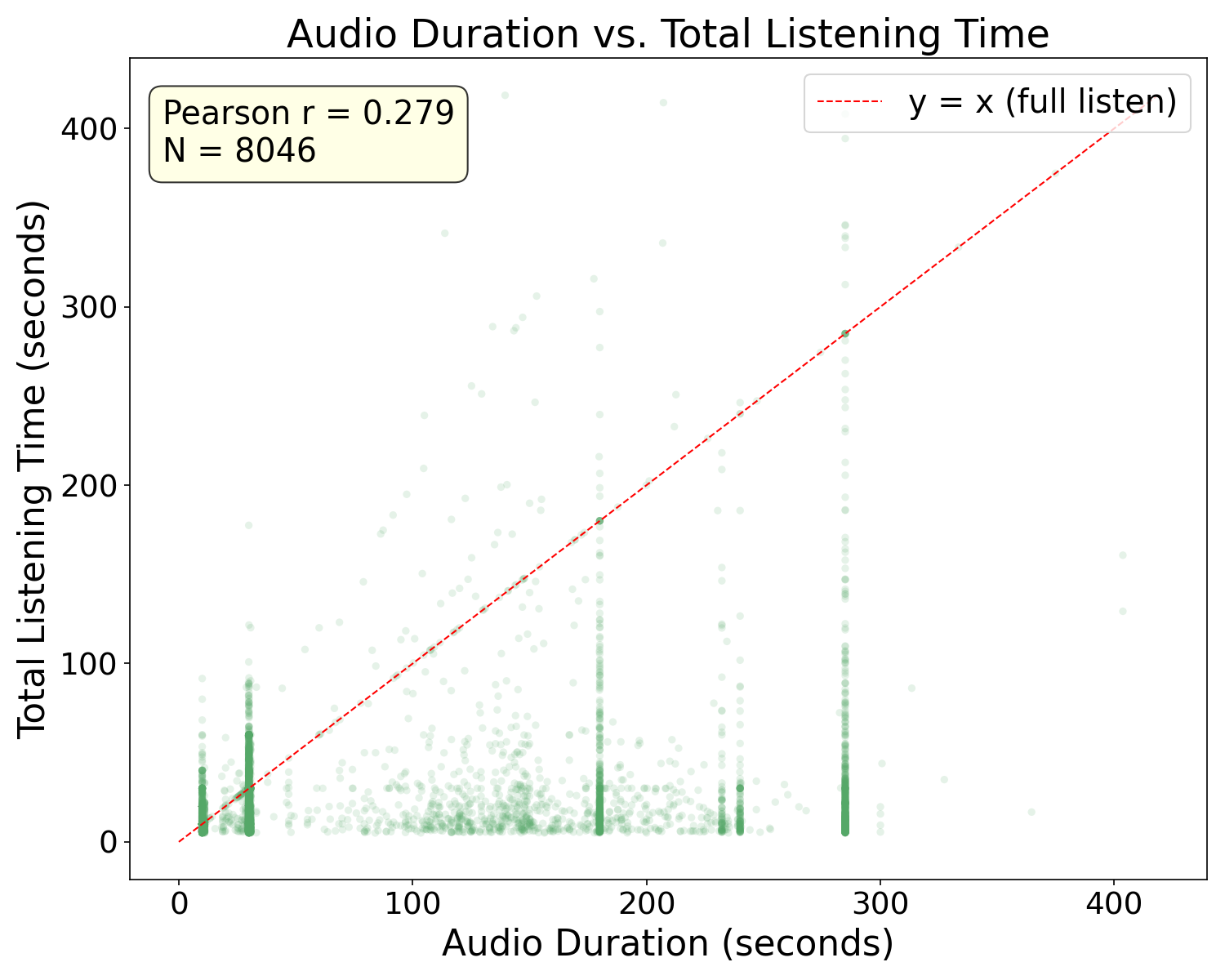
  }
  \caption{Correlation between audio duration (x-axis) and total listening
    time (y-axis), computed on $N=8046$ filtered samples. The dashed diagonal
    denotes $y=x$ (full listening).}
  \label{fig:duration_vs_listening_time_scatter}
\end{figure}

Figure~\ref{fig:duration_vs_listening_time_scatter} examines the relationship
between audio duration and total listening time on the same filtered set ($N=80
  46$). We observe a mild positive correlation (Pearson $r=0.279$), indicating
that longer audios are associated with longer listening on average, but the dependence
is limited.

Notably, most points lie below the diagonal $y=x$, meaning annotators often make
decisions before fully listening to the entire clip. Meanwhile, the broad spread
at fixed duration shows substantial variability in listening behavior across examples.
Together, these findings indicate that annotators adaptively allocate listening
effort by sample difficulty rather than simply following clip length.

\subsection{Annotator Agreement in CMI-Pref}
\label{app:annotator-agreement}
\begin{table}[h]
  \centering
  \small
  \caption{Inter-annotator agreement on overlapping votes, computed over 644 vote
    pairs from 492 comparisons with multiple annotations.}
  \label{tab:overlapping_votes} \resizebox{\linewidth}{!}{%
    \begin{tabular}{lcc}
      \toprule \textbf{Metric} & \textbf{Instruction Following} & \textbf{Music Quality} \\
      \midrule Agreement Count & 445                            & 466                    \\
      Disagreement Count       & 199                            & 178                    \\
      Agreement Rate           & 0.691                          & 0.724                  \\
      Krippendorff’s $\alpha$  & 0.382                          & 0.447                  \\
      \bottomrule
    \end{tabular}
  }
\end{table}

\begin{table}[h]
  \centering
  \small
  \caption{Reliability of CMI-Pref-test, computed by comparing three annotators'
    re-annotations of all 500 test samples with the original test labels.}
  \label{tab:test_reannotation} \resizebox{\linewidth}{!}{%
    \begin{tabular}{lcc}
      \toprule \textbf{Metric} & \textbf{Alignment Preference} & \textbf{Musicality Preference} \\
      \midrule Agreement Rate  & 0.750                         & 0.752                          \\
      Krippendorff's $\alpha$  & 0.500                         & 0.504                          \\
      Wilson 95\% CI           & (70.9\%, 78.5\%)              & (71.1\%, 78.7\%)               \\
      \bottomrule
    \end{tabular}
  }
\end{table}

We analyze inter-annotator consistency for comparisons that received multiple
independent votes. A total of 492 comparison pairs were annotated by two or more
annotators, producing 1,056 votes. For each comparison, we form all pairwise
combinations of its votes, resulting in 644 vote pairs.

We measure agreement using \textit{agreement rate} and \textit{Krippendorff's
  alpha}, computed separately for alignment and musicality. For the training split,
Table~\ref{tab:overlapping_votes} reports duplicated-vote agreement on overlapping
annotations. Music quality judgments exhibit slightly higher consistency than
instruction following, as they are more intuitive and place lower demands on
specialized musical knowledge. After correcting for chance agreement, Krippendorff's
alpha indicates a \textit{moderate} level of true agreement, which we consider
reasonable given the inherent subjectivity of music evaluation.

To directly assess the reliability of CMI-Pref-test, we additionally re-annotated
all 500 test samples with 4 annotators. As shown in Table~\ref{tab:test_reannotation}, agreement
between the re-annotations and the original test labels is substantially higher
than the overlap statistics in Table~\ref{tab:overlapping_votes}. For musicality
preference, we obtain $\alpha=0.504$ and agreement $=75.2\%$; for alignment
preference, we obtain $\alpha=0.500$ and agreement $=75.0\%$. We also examined
whether model performance is unrealistically high relative to human consistency.
The strongest models are generally close to this re-annotation agreement level:
CMI-RM reaches 74\% accuracy for alignment and 78\% for musicality, while
SongEval-RM reaches 72.4\% for musicality. This supports the validity of the
benchmark rather than suggesting overfitting to noisy labels.

\subsection{Confidence Scores of Human Annotation}
\begin{figure}[htb]
  \centering
  \includegraphics[width=.96\columnwidth]{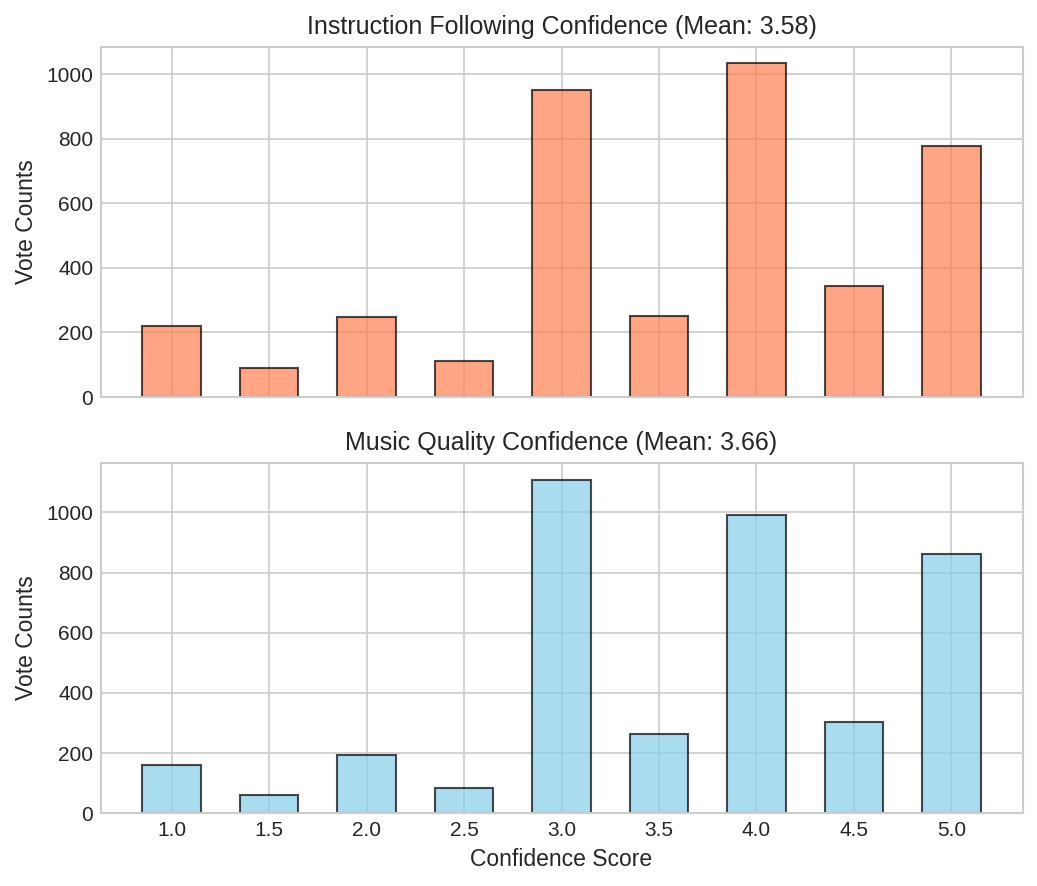}
  \caption{Distribution of annotators' confidence scores for instruction following
    and music quality.}
  \label{fig:human_confidence_distribution}
\end{figure}

We analyze the distribution of annotators’ confidence scores (Fig.~\ref{fig:human_confidence_distribution}).
Overall, most votes are associated with relatively high confidence. The
average confidence for music quality is slightly higher than that for
instruction following, consistent with earlier observations that music quality
judgments are more intuitive.

\begin{table}[ht]
  \centering
  \small
  \caption{Agreement between instruction-following and music-quality
    preferences, and the corresponding average confidence.}
  \label{tab:if_mq_agreement}
  \begin{tabular}{lc}
    \toprule \textbf{Metric}              & \textbf{Value}        \\
    \midrule \#Votes with same preference & 3309                  \\
    \#Votes with different preference     & 718                   \\
    Agreement rate                        & 0.822                 \\
    Avg. confidence (same Pref.)          & IF: 3.674,\ MQ: 3.734 \\
    Avg. confidence (diff. Pref.)         & IF: 3.171,\ MQ: 3.311 \\
    \bottomrule
  \end{tabular}
\end{table}

We further examine the relationship between agreement across instruction following
and music quality and annotator confidence (Table~\ref{tab:if_mq_agreement}).
Overall, the two dimensions show a high agreement rate, and votes have higher
average confidence when they agree. This suggests that disagreements between instruction
following and music quality often correspond to more ambiguous or trade-off cases,
where annotators are less certain about the overall preference.

\textbf{Alignment among heads:}
We observed a high consistency between musicality and instruction-music alignment
preferences. In pseudo-labels, the agreement is 91\% between musicality and
instruction alignment. In human data, this agreement is lower (81\%), but
conflicts typically occur when annotators report high confidence in
instruction alignment despite lower musicality, highlighting the necessity of evaluating
these dimensions separately.

\section{Pseudo-label Generation for CMI-Pseudo}
\label{app:pseudo-label-generation}

\subsection{Label Acquisition Protocol}
To ensure the reliability of our pseudo-labels, we address the well-documented
phenomenon of positional bias in Large Language Models (LLMs), where the model's
preference is influenced by the presentation order of the options rather than
the content quality alone.

Our protocol for collecting pseudo-labels adopts a \textit{Position-Consistency}
strategy. For a given pair of audio samples $(A, B)$ generated from the same
prompt $P$, we conduct a bidirectional assessment:
\begin{enumerate}
  \item \textbf{Forward Pass:} We query the model with the sequence $(A, B)$
        to obtain preference $L_{fwd}$.

  \item \textbf{Reverse Pass:} We swap the positions to $(B, A)$ and query the
        model again to obtain preference $L_{rev}$.
\end{enumerate}
A pseudo-label is considered valid and retained only if the judgment is
invariant to position—that is, the model prefers the same underlying audio clip
in both the forward and reverse passes ($L_{fwd}= L_{rev}$). Comparisons
yielding conflicting results or inconsistent ties are discarded as
hallucinations or uncertain boundaries.

\subsection{Bias Analysis and Dataset Statistics}
We initially sampled 129,545 pairwise comparisons from our generated audio
pool. As demonstrated in Table \ref{tab:bias}, we observe significant
distributional shifts when the presentation order is swapped, confirming the presence
of positional bias. For instance, in the Musicality metric, the win rate for
Candidate A fluctuates from 51.96\% in the original configuration to 59.27\%
in the reversed configuration.

By applying our consistency filter, we retain only the high-confidence labels,
resulting in 114,694 valid musicality labels and 117,828 valid alignment
labels. As shown in the ``Agreed" rows of Table \ref{tab:bias}, the resulting distribution
stabilizes, effectively mitigating the variance introduced by the model's
sensitivity to input order. For the final dataset construction, we retained
the \textbf{intersection} of these valid subsets to ensure quality across all dimensions,
yielding approximately 110k pairs.

\begin{table}[!htbp]
  \centering
  \small %
  \caption{Distribution of pseudo-labels across varying query configurations. ``Original"
    and ``Reversed" denote the presentation order of the audio pair. ``Agreed" represents
    the subset of labels where the model's preference remained consistent across
    both permutations. The discrepancy between Original and Reversed highlights
    positional bias, which is rectified in the Agreed set.}
  \label{tab:bias}
  \begin{tabular}{lccc}
    \toprule \textbf{Configuration}         & \textbf{Win A (\%)} & \textbf{Win B (\%)} & \textbf{Tie (\%)} \\
    \midrule \textit{Musicality}            &                     &                     &                   \\
    \hspace{3mm} Original $(A, B)$          & 51.96               & 40.96               & 7.08              \\
    \hspace{3mm} Reversed $(B, A)$          & 59.27               & 33.48               & 7.25              \\
    \hspace{3mm} \textbf{Agreed (Filtered)} & \textbf{57.80}      & \textbf{36.97}      & \textbf{5.22}     \\
    \midrule \textit{Alignment}             &                     &                     &                   \\
    \hspace{3mm} Original $(A, B)$          & 33.65               & 43.71               & 22.64             \\
    \hspace{3mm} Reversed $(B, A)$          & 38.75               & 38.84               & 22.77             \\
    \hspace{3mm} \textbf{Agreed (Filtered)} & \textbf{36.05}      & \textbf{41.23}      & \textbf{22.73}    \\
    \bottomrule
  \end{tabular}
\end{table}

The ``Original" and ``Reversed" distributions should theoretically be identical
if the evaluator were perfectly unbiased. The observed divergence necessitates
our rigorous filtering approach. We did not shuffle the comparison pairs during
analysis, therefore its normal that the labels A, B are uneven.

\section{Human Annotation Details}
\label{app:human-annotation}
\subsection{Annotation Protocol for CMI-Pref}

\subsubsection{Core Objectives}
The annotation process is decomposed into three distinct components to ensure a
multi-dimensional evaluation of the generated music:
\begin{enumerate}
  \item \textbf{Preference Label (A/B):} A forced-choice selection between two
        candidates.

  \item \textbf{Confidence Score (1--5):} A quantitative measure of the
        annotator's certainty, grounded in constraint satisfaction (for alignment)
        or quality delta (for musicality).

  \item \textbf{Free-text Feedback:} Qualitative justifications focusing on
        fine-grained details that discrete labels cannot capture.
\end{enumerate}

\subsubsection{General Principles}
\begin{itemize}
  \item \textbf{Instruction-First:} Annotators must strictly evaluate the
        \textit{instruction/prompt} before listening to avoid post-hoc rationalization.

  \item \textbf{Holistic and Granular Review:} Each sample is evaluated for
        overall coherence as well as specific details (instrumentation, emotion, structure,
        and audio fidelity).

  \item \textbf{Dimensional Isolation:} Annotators are instructed to decouple
        \textbf{Alignment} (adherence to the prompt) from \textbf{Musicality} (aesthetic
        quality and production value). A sample may win in alignment while losing in
        musicality.
\end{itemize}

\subsubsection{Q1: Textual Music Alignment Preference}
Annotators identify which sample better follows the elements specified in the
instruction, regardless of aesthetic appeal. Key alignment dimensions include:
\begin{itemize}
  \item \textbf{Instrumentation:} Specific instruments (e.g., piano solo,
        guitar riff).

  \item \textbf{Mood/Atmosphere:} Emotional valence (e.g., melancholy, upbeat,
        tense).

  \item \textbf{Genre/Style/Era:} Stylistic markers (e.g., Lo-fi, Baroque, 80s
        synth).

  \item \textbf{Rhythm/Tempo:} Temporal characteristics (e.g., driving drum beat,
        groovy, energetic).
\end{itemize}

\textbf{Confidence Calibration (1--5):}
\begin{itemize}
  \item \textbf{5 (Very Certain):} Clear binary distinction; one satisfies all
        key constraints while the other fails or collapses.

  \item \textbf{3 (Moderate):} Default choice; a perceptible lean toward one candidate
        without overwhelming dominance.

  \item \textbf{1 (Uncertain):} Highly ambiguous prompts or both samples are
        indistinguishable in their failure/success.
\end{itemize}

\subsubsection{Q2: Musicality Preference}
Annotators evaluate which sample sounds more like a ``finished, natural, and
professional" musical work, independent of the prompt.
\begin{itemize}
  \item \textbf{Key Criteria:} Melodic memorability, structural progression,
        rhythmic stability, and production clarity (lack of distortion/artifacts).

  \item \textbf{Confidence:} Reflects the perceived ``quality gap" between candidates.
\end{itemize}

\subsubsection{Q3: Feedback Guidelines}
Feedback should consist of 1--3 concise sentences focusing on ``audible evidence."
Annotators are encouraged to use specific timestamps and avoid vague descriptors.
\begin{itemize}
  \item \textbf{Positive Justification:} ``Sample A aligns better due to the
        presence of the specified saxophone; the melody is more distinct."

  \item \textbf{Negative Evidence:} ``Sample B suffers from rhythmic
        instability at 0:20 and harsh high-frequency distortion."
\end{itemize}

\subsubsection{Glossary for Annotators}
\begin{table}[h]
  \centering
  \small
  \caption{Taxonomy of musical attributes used in the annotation process.}

  \resizebox{\linewidth}{!}{%

    \begin{tabular}{lll}
      \hline
      \textbf{Dimension}     & \textbf{Positive Descriptors}      & \textbf{Negative Descriptors}    \\
      \hline
      \textbf{Melody}        & Catchy, Memorable, Distinct        & Repetitive, Generic, Wandering   \\
      \textbf{Structure}     & Coherent, Progression, Build-up    & Disjointed, Abrupt, Random Loops \\
      \textbf{Rhythm}        & Groovy, Steady, Driving            & Off-beat, Unstable, Chaotic      \\
      \textbf{Audio Quality} & Clean mix, Balanced, High-fidelity & Harsh, Muddy, Distorted/Clipping \\
      \textbf{Vocals}        & Natural, Clear articulation        & Robotic, Slurred, Artifacts      \\
      \hline
    \end{tabular}
  }
\end{table}
\subsection{Annotation Platform}
\begin{figure}[h]
  \vskip 0.2in
  \begin{center}
    \centerline{\includegraphics[width=.96\columnwidth]{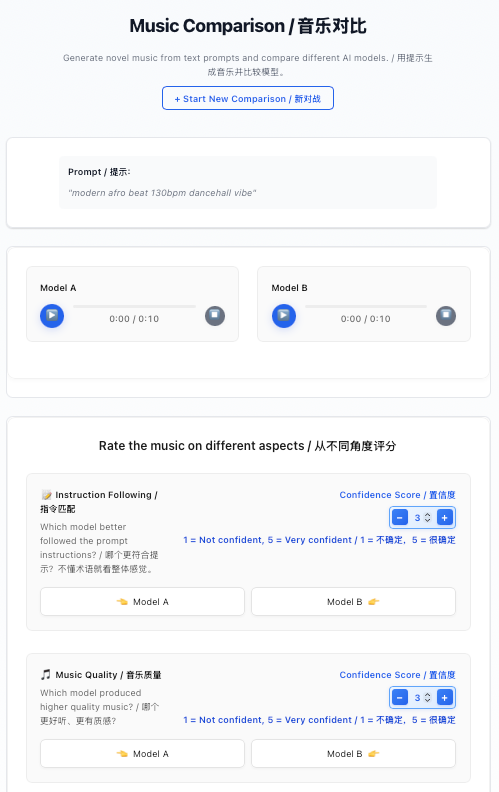}}
    \caption{ Platform of human annotation. }
    \label{fig:arena-web}
  \end{center}
\end{figure}

Our annotation platform is illustrated in \autoref{fig:arena-web}. It provides
a unified interface for pairwise audio comparison, confidence scoring, and
free-text feedback collection. Detailed usage instructions are provided in the
README included in the submitted supplementary materials.

\section{Limitations}

Our reward model is designed as a practical \textbf{baseline evaluator} for
compositional music preference modeling, rather than a complete solution for
reward optimization in downstream RL. In particular, we do not claim that a
single reward model prevents reward hacking by itself.

We observe \textbf{over-correlation between heads}. In human labels, musicality
and alignment preferences agree on around \textbf{82\%} of votes; in model
predictions, the SRCC between alignment and musicality scores is around
\textbf{0.853}. This correlation helps transfer, but also suggests incomplete
disentanglement when musicality and alignment conflict.

We also observe \textbf{bias by genre and language}. Musicality scores vary more
across genres than alignment scores, with highly structured acoustic genres (classical, jazz, pop)
tending to receive higher scores than electronic or experimental genres.
However, on pairwise preference evaluation, this genre effect is less severe,
since lower-scoring genres still show near-average accuracy. We also observe
language imbalance: alignment performance is stronger on English than on
low-resource languages.

\section{Ablation Study on Pseudo-Labeled Data}
\subsection{Distribution Shift and the Effectiveness of Label Smoothing}
\label{Append:Label Smoothing}

We first pretrain the reward head on pseudo-labeled data until convergence ($\sim$14k
steps), and then fine-tune it on real preference data. Although pretraining on
AI feedback already yields reasonably high accuracy on the real test set (\textbf{71.0\%}
on Musicality and \textbf{71.8\%} on Alignment), it provides only \emph{limited}
additional benefit after downstream fine-tuning: pseudo-pretraining without
smoothing reaches \textbf{75.2\%}/\textbf{71.0\%} after fine-tuning, comparable
to directly fine-tuning from scratch (\textbf{75.0\%}/\textbf{69.3\%}).

\begin{figure}[t]
  \centering
  \includegraphics[width=\linewidth]{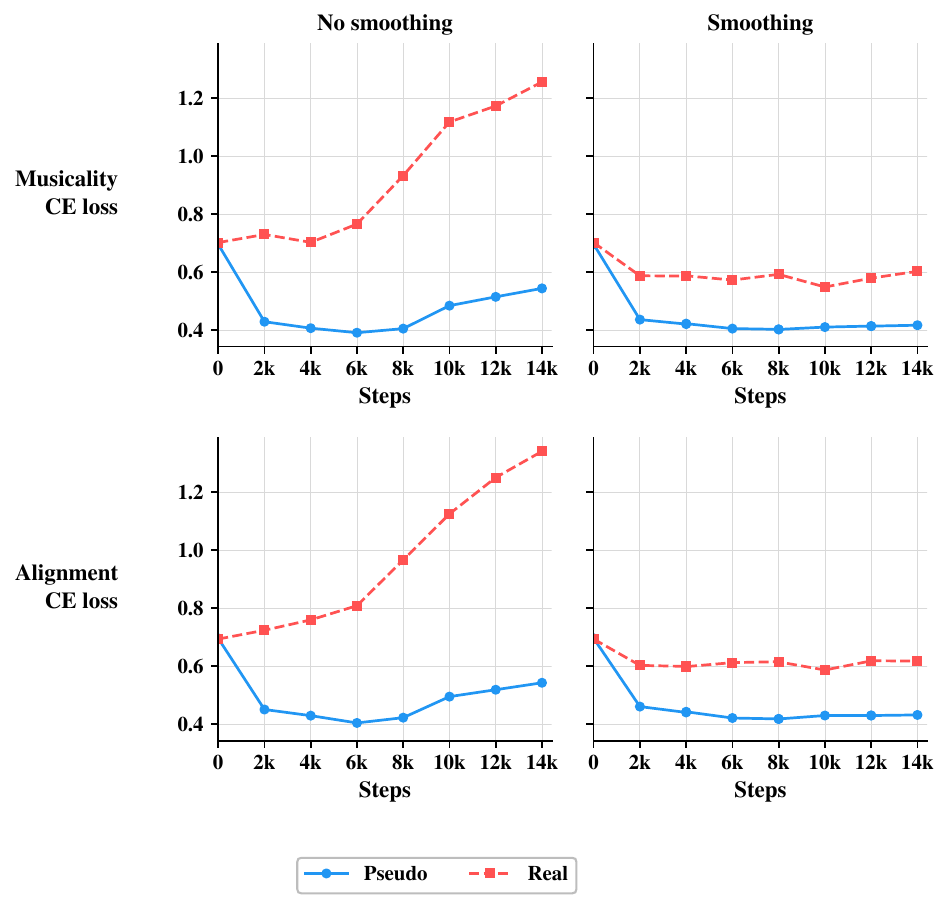}
  \caption{Cross-entropy loss of checkpoints trained on CMI-Pseudo. \textit{Pseudo}
    indicates metrics on the CMI-Pseudo validation set, and \textit{Real}
    denotes results on the test set of CMI-Pref.}
  \label{fig:ce_trend}
\end{figure}

\begin{figure}[t]
  \centering
  \includegraphics[width=\linewidth]{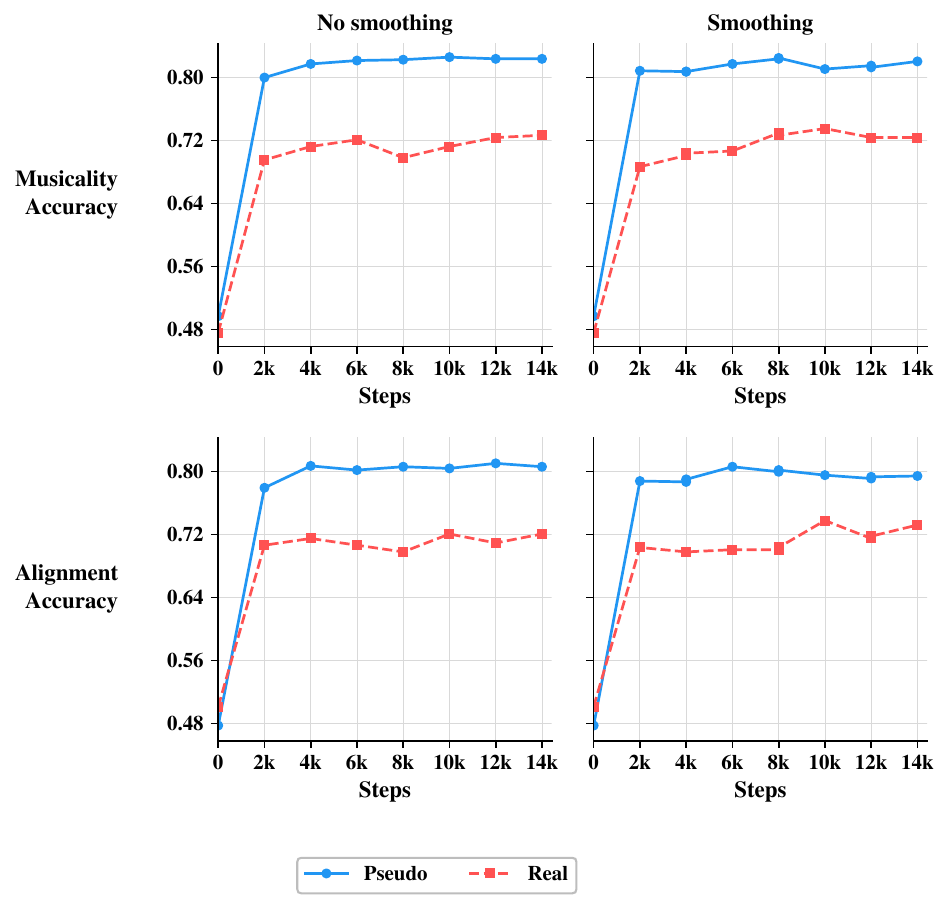}
  \caption{Accuracy of checkpoints trained on CMI-Pseudo. \textit{Pseudo}
    indicates metrics on the CMI-Pseudo validation set, and \textit{Real}
    denotes results on the test set of CMI-Pref.}
  \label{fig:acc_trend}
\end{figure}

We attribute this to a distribution shift between pseudo-labeled and real data,
which manifests as over-confident probabilities. Specifically, we track the
cross-entropy loss on the real test set, We measure binary cross-entropy (CE)
\[
  \mathcal{L}_{\mathrm{CE}}= - \bigl(y\log p_{\theta}(A \succ B) + (1-y)\log(1-
  p_{\theta}(A \succ B))\bigr),
\]
and observe that it can exceed \textbf{1.2} initially, while a random guess baseline
yields $-\ln(0.5)=0.693$ for a binary task. This indicates that the model can
be confidently wrong under the shift, which is penalized heavily by cross-entropy
even when accuracy changes only slightly. Figure~\ref{fig:ce_trend} shows that
without label smoothing, the CE loss on the pseudo validation set remains stable,
whereas the CE loss evaluated on real data increases substantially. In contrast,
the corresponding accuracy curves in Figure~\ref{fig:acc_trend} show much
smaller differences, suggesting the model produces over-confident decision boundaries.

To mitigate this over-confidence issue, during pseudo training, we apply label
smoothing with parameter $\varepsilon$:
\[
  \tilde{y}=(1-\varepsilon)y+\varepsilon/2,
\]
which replaces hard targets $y\in\{0,1\}$ with softened targets
$\tilde{y}\in(0,1)$. We also only train the model 1.5 epochs. In our
experiments, we set $\varepsilon=0.2$, i.e., $(1,0)\rightarrow(0.9,0.1)$. This
simple change yields clear downstream gains: pseudo-pretraining \emph{with} smoothing
achieves \textbf{77.8\%} on Musicality and \textbf{74.0\%} on Alignment after fine-tuning,
outperforming both direct fine-tuning and pseudo-pretraining without smoothing.

Label smoothing makes checkpoint selection more robust. With smoothing,
\textsc{CMI-Pref}-test accuracy stays around $75.3\pm1\%$ from 2k to 20k steps,
while training without smoothing is more sensitive to stopping points and yields
worse downstream transfer. However, this robustness is not uniform across
datasets: Music Arena starts declining after around 6k steps, and PAM is more
sensitive to over-training. This is possible due to dataset-specific inconsistencies.
Therefore, we still maintain a relative early stopping point (1.5 epochs) to prevent overfitting to specific patterns.

\begin{table}[t]
  \centering
  \small
  \caption{Real-test reward-head accuracy (\%) under different training
    pipelines.}
  \label{tab:smoothing_downstream} \resizebox{\linewidth}{!}{
    \begin{tabular}{lcc}
      \hline
      \textbf{Training pipeline}                        & \textbf{Musicality Acc.} & \textbf{Alignment Acc.} \\
      \hline
      Pseudo pretrain (before FT)                       & 71.0\%                   & 71.8\%                  \\
      Pseudo pretrain $\rightarrow$ FT (no smoothing)   & 75.2\%                   & 71.0\%                  \\
      Direct FT (from scratch)                          & 75.0\%                   & 69.3\%                  \\
      Pseudo pretrain $\rightarrow$ FT (with smoothing) & \textbf{77.8\%}          & \textbf{74.0\%}         \\
      \hline
    \end{tabular}}
\end{table}

\subsection{Ablation on Pseudo-Label Data Size}
\label{subsec:pseudo_size}

We ablate the amount of pseudo-labeled data used for pre-training. To simplify
the evaluation protocol, we fine-tune only on \textsc{CMI-Pref}. Unless
otherwise noted, label smoothing is enabled during pseudo pre-training.

\paragraph{Metrics.}
We report two groups of metrics. \textbf{Transfer before fine-tuning:} we
evaluate the pretrained model on the \textsc{CMI-Pref} test set and report (i)
\textbf{Pref-Test Acc.}, defined as the average reward-head accuracy over the Musicality
and Alignment heads, and (ii) \textbf{Pref-Test CE}, defined as the
corresponding average cross-entropy. \textbf{Downstream after fine-tuning:}
after fine-tuning on \textsc{CMI-Pref}, we report the head accuracies: \textbf{Mus-Acc}
(Musicality head accuracy) and \textbf{Align-Acc} (Alignment head accuracy).

\paragraph{Setup.}
The full pseudo-labeled dataset contains \textbf{110k} examples. We vary the
pseudo data size in
$\{4\text{k}, 8\text{k}, 16\text{k}, 32\text{k}, 64\text{k}, 110\text{k}\}$. We
primarily compare models at a fixed \textbf{epoch progress} during pseudo pre-training
(Table~\ref{tab:ablation_epoch}), which provides a natural normalization across
different dataset sizes. For completeness, we also include a fixed-step comparison
in Table~\ref{tab:ablation_step}.

\paragraph{Results.}
Under the same-epoch protocol (Table~\ref{tab:ablation_epoch}), increasing the
pseudo data size generally improves both transfer (Pref-Test Acc/CE) and downstream
performance. Moving from small pseudo sets (4k--16k) to \textbf{32k--64k}
yields noticeably stronger results, while performance saturates from \textbf{64k}
to \textbf{110k}. Notably, \textbf{64k} achieves performance comparable to
using the full pseudo-labeled set under the same-epoch protocol, suggesting diminishing
returns beyond this scale. Therefore, we adopt \textbf{64k} pseudo-labeled
examples as a compute-efficient operating point in subsequent experiments. Since
directly finetuning on the 3.5k CMI-Pref yields Pref-Text Acc of 0.7215, we validate
our rule-of-thumb that Psuedo data should be more than 10 times compared to real
data. The ablation on epoch vs step also demonstrates that epoch is a more
natural normalization in this finetuning setting.

\begin{table}[t]
  \centering
  \small
  \caption{Pseudo data size ablation under the fixed epoch-progress protocol (same-epoch
    comparison).}
  \label{tab:ablation_epoch} \resizebox{\columnwidth}{!}{
    \begin{tabular}{lcccc}
      \toprule Pseudo Size & Pref-Test Acc.$\uparrow$ & Pref-Test CE$\downarrow$ & Mus-Acc$\uparrow$ & Align-Acc$\uparrow$ \\
      \midrule 4k          & 0.582                    & 0.674                    & 0.726             & 0.664               \\
      8k                   & 0.640                    & 0.656                    & 0.750             & 0.692               \\
      16k                  & 0.605                    & 0.655                    & 0.754             & 0.700               \\
      32k                  & 0.646                    & 0.627                    & 0.782             & 0.700               \\
      64k                  & 0.711                    & 0.593                    & 0.778             & 0.740               \\
      110k                 & 0.720                    & 0.559                    & 0.761             & 0.735               \\
      \bottomrule
    \end{tabular}}
\end{table}

\begin{table}[H]
  \centering
  \small
  \caption{Pseudo data size ablation under the fixed-step protocol (same-step comparison).}
  \label{tab:ablation_step} \resizebox{\columnwidth}{!}{
    \begin{tabular}{lcccc}
      \toprule Pseudo Size & Pref-Test Acc.$\uparrow$ & Pref-Test CE$\downarrow$ & Mus-Acc$\uparrow$ & Align-Acc$\uparrow$ \\
      \midrule 4k          & 0.588                    & 0.724                    & 0.742             & 0.696               \\
      8k                   & 0.623                    & 0.685                    & 0.768             & 0.694               \\
      16k                  & 0.646                    & 0.636                    & 0.760             & 0.682               \\
      32k                  & 0.566                    & 0.725                    & 0.752             & 0.710               \\
      64k                  & 0.711                    & 0.593                    & 0.778             & 0.740               \\
      110k                 & 0.712                    & 0.570                    & 0.762             & 0.714               \\
      \bottomrule
    \end{tabular}}
\end{table}

\subsection{Ablation on Mixed Pseudo-Label Sources}

We also test whether mixing pseudo labels from multiple teachers is more
effective than Qwen-only supervision with label smoothing. The mixed setting
uses \textbf{70k} Qwen3-Omni labels and \textbf{40k} Gemini labels.
Table~\ref{tab:mixed_pseudo_ablation} shows that source mixing provides only
modest and inconsistent gains. Mixing alone improves PAM and
\textsc{CMI-Pref}, but slightly hurts Music Arena. Combining smoothing and
mixing gives the best \textsc{CMI-Pref} and Music Arena results, but degrades
PAM. Overall, label smoothing remains the more reliable regularizer, while
multi-source distillation mainly changes the teacher boundary.

\begin{table}[t]
  \centering
  \small
  \caption{Aggregated results under different pseudo-label source designs.}
  \label{tab:mixed_pseudo_ablation}
  \resizebox{\columnwidth}{!}{
    \begin{tabular}{lcccc}
      \toprule
      \textbf{Pseudo-label setup} & \makecell{\textbf{PAM Mean}                         \\\textbf{SRCC}} & \makecell{\textbf{MusicEval}\\\textbf{SRCC}} & \makecell{\textbf{Music Arena}\\\textbf{Acc}} & \makecell{\textbf{CMI-Pref}\\\textbf{Mean Acc}} \\
      \midrule
      Smooth (Qwen only)          & 0.529                       & 0.742 & 0.701 & 0.756 \\
      Mix (70k Qwen + 40k Gemini) & 0.550                       & 0.761 & 0.699 & 0.768 \\
      Smooth + Mix                & 0.450                       & 0.765 & 0.707 & 0.775 \\
      \bottomrule
    \end{tabular}
  }
\end{table}

\section{Reward Model}

\subsection{Model Size Scaling}

Our main configuration keeps pretrained encoders frozen and trains only the
reward head. To study parameter efficiency, we vary hidden size, attention
heads, and Transformer depth to obtain three reward-head sizes: \textbf{small}
(8.3M), \textbf{base} (38M), and \textbf{large} (102M). All variants are
pretrained on \textsc{CMI-Pref-Pseudo} and then fine-tuned on
\textsc{CMI-Pref} and MusicEval.

Table~\ref{tab:model_size_ablation} shows that smaller models can still work
well: the 8.3M variant remains competitive on PAM, MusicEval, and
\textsc{CMI-Pref}, and mainly underperforms on Music Arena. The 102M variant
achieves the strongest overall results, especially on MusicEval and
\textsc{CMI-Pref}, indicating better multi-task transfer with larger heads.

\begin{table}[t]
  \centering
  \small
  \caption{Model size scaling of the trainable reward head.}
  \label{tab:model_size_ablation}
  \setlength{\tabcolsep}{4pt}
  \resizebox{\columnwidth}{!}{
    \begin{tabular}{lccccc}
      \toprule
      \textbf{Variant} & \makecell[c]{\textbf{Trainable}                                 \\\textbf{Params}} & \makecell[c]{\textbf{PAM}\\\textbf{Mean SRCC}} & \makecell[c]{\textbf{MusicEval}\\\textbf{SRCC}} & \makecell[c]{\textbf{MusicArena}\\\textbf{Acc}} & \makecell[c]{\textbf{CMI-Pref}\\\textbf{Mean Acc}} \\
      \midrule
      Small            & 8.3M                            & 0.540 & 0.816 & 0.680 & 0.744 \\
      Base             & 38M                             & 0.578 & 0.811 & 0.729 & 0.746 \\
      Large            & 102M                            & 0.589 & 0.848 & 0.719 & 0.765 \\
      \bottomrule
    \end{tabular}
  }
\end{table}

\subsection{Text Encoder Ablation with Flan-T5}

We replace the default text encoder with \textbf{Flan-T5-Large} to test whether
stronger text modeling alone improves compositional reward prediction. For this
ablation, we use the \textit{small} reward-head setting. We report changes in
accuracy relative to the baseline fine-tuned on \textsc{CMI-Pref}.

As shown in Table~\ref{tab:t5_ablation}, Flan-T5 provides a gain on one
text-only Music Arena subset, is neutral on CMI text+lyrics, and degrades
consistently when audio conditioning is present. This suggests that the current
bottleneck is cross-modal audio-text fusion rather than text encoder capacity.

\begin{table}[t]
  \centering
  \small
  \caption{Accuracy change from replacing the default text encoder with
    Flan-T5-Large. T/L/A denote text, lyrics, and audio conditions.}
  \label{tab:t5_ablation}
  \resizebox{\columnwidth}{!}{
    \begin{tabular}{lcccccc}
      \toprule
      \textbf{Metric} & \textbf{Arena T} & \textbf{Arena T+L} & \textbf{CMI T} & \textbf{CMI T+L} & \textbf{CMI T+L+A} & \textbf{CMI T+A} \\
      \midrule
      $\Delta$ Acc.   & +2.7\%           & -1.4\%             & -4.0\%         & 0.0\%            & -7.2\%             & -6.8\%           \\
      \bottomrule
    \end{tabular}
  }
\end{table}

\subsection{Ablation Study on Mapping Functions}
We investigate the impact of different mapping functions used to transform the
predicted preference scores for MOS regression. We compare four settings:
\begin{itemize}
  \item \textbf{None}: No transformation is applied.

  \item \textbf{Tanh}: The method adopted in our main experiments, using a
        Tanh activation to bound the output.

  \item \textbf{Linear}: A linear projection without Tanh activation.

  \item \textbf{Ordinal}: We employ an ordinal regression objective with learnable
        classification margins during training. Note that for evaluation, we use the
        model's raw scalar output to compute metrics, bypassing the learned
        margins.
\end{itemize}
To ensure a fair comparison, all models are initialized from the CMI-Pref-Pseudo
pretraining checkpoint (2,000 steps) and finetuned on CMI-Pref and MusicEval. As
shown in Table~\ref{tab:loss_ablation}, we observe that the choice of mapping
function does not yield significant performance differences.
\begin{table}[H]
  \small
  \caption{Aggregated results on mapping ablations}
  \label{tab:loss_ablation} \resizebox{\columnwidth}{!}{
    \begin{tabular}{lcccc} %
      \toprule Datasets  & PAM       & MusicEval & Music Arena & CMI-Pref \\
      Metrics $\uparrow$ & Mean SRCC & SRCC      & ACC         & Mean ACC \\
      \midrule None      & 0.5069    & 0.4659    & 73.73\%     & 75.3\%   \\ %
      Tanh               & 0.6024    & 0.4702    & 73.8\%      & 74.8\%   \\
      Linear             & 0.4595    & 0.4691    & 73.58\%     & 74.40\%  \\
      Ordinal            & 0.6079    & 0.4665    & 72.31\%     & 75.80\%  \\
      \hline
    \end{tabular}
  }
\end{table}

\subsection{The Effect of Duration on Model Inference}

The variable-length nature of audio makes it important to analyze the effect
of duration at inference time. Scoring an entire waveform can be memory-intensive,
while restricting inference to short windows may reduce compute but risk
missing long-range musical structure and text--music relationships. We
evaluate three inference strategies using the same reward model (w/ f.t.: CMI
+ MusicEval):
\begin{enumerate}
  \item \textit{first 10}: Use only the first 10 seconds of audio to infer the
        Musicality and Alignment scores.

  \item \textit{mean 10}: Split the audio into non-overlapping 10-second chunks
        (hop size 10 seconds), infer scores for each chunk, and take the average as
        the final score.

  \item \textit{first 120}: Match the training setting by extracting MuQ embeddings
        from the first 120 seconds (concatenating four 30-second segments), and
        then infer scores using the learned weights.
\end{enumerate}
Among these, \textit{mean 10} is the most compute-intensive. With a maximum duration
of 120 seconds and batched inference, its memory overhead is only slightly
larger than \textit{first 120}, but it requires more forward passes.

Table~\ref{tab:duration_ablation} reports accuracy on CMI-RewardBench subsets
across duration bins. We do not report metrics on PAM and MusicEval due to their
short duration( $\le60$ for MusicEval and $\le$30 for PAM). On Music Arena,
\textit{first 120} achieves the best performance in most bins and shows clear gains
over \textit{first 10}, suggesting that short excerpts often under-represent relevant
musical content. \textit{mean 10} attains comparable accuracy to \textit{first
  120} in several cases, indicating that aggregating local judgments can approximate
longer-context inference. On CMI-Pref, the same trend generally holds but is less
pronounced: \textit{first 10} remains competitive, especially in shorter-duration
bins, implying that some preference decisions can be resolved from early content.

\begin{table}[H]
  \centering
  \caption{Accuracy (\%) by audio duration across inference variants. Highest per
    bin is \textbf{bold}.}
  \label{tab:duration_ablation} \footnotesize
  \setlength{\tabcolsep}{3.5pt}
  \begin{tabular}{llrrrr}
    \toprule                 &                    & \multicolumn{4}{c}{Duration (s), Acc (\%)}                                                 \\
    \cmidrule(l){3-6} Metric & Variant            & {[}10,30{)}                                & {[}30,60{)}   & {[}60,120{)}  & {[}120,500{]} \\
    \midrule \multirow{4}{*}{\shortstack[l]{Music                                                                                              \\Arena}}         & \textit{n}         & \textit{119}                              & \textit{88}   & \textit{531}  & \textit{600}  \\
                             & \textit{first 10}  & 72.3                                       & \textbf{53.4} & 64.2          & 59.5          \\
                             & \textit{mean 10}   & \textbf{75.6}                              & 48.9          & 75.7          & 67.0          \\
                             & \textit{first 120} & \textbf{75.6}                              & \textbf{53.4} & \textbf{81.1} & \textbf{67.3} \\
    \midrule \multirow{4}{*}{\shortstack[l]{CMI-Pref                                                                                           \\Musicality}} & \textit{n}         & \textit{196}                              & \textit{77}   & \textit{73}   & \textit{154}  \\
                             & \textit{first 10}  & 77.6                                       & 71.4          & 87.7          & \textbf{76.0} \\
                             & \textit{mean 10}   & \textbf{78.1}                              & \textbf{76.6} & 89.0          & 74.7          \\
                             & \textit{first 120} & 77.6                                       & \textbf{76.6} & \textbf{90.4} & 73.7          \\
    \midrule \multirow{4}{*}{\shortstack[l]{CMI-Pref                                                                                           \\Alignment}}  & \textit{n}         & \textit{196}                              & \textit{77}   & \textit{73}   & \textit{154}  \\
                             & \textit{first 10}  & 68.4                                       & 74.0          & \textbf{83.6} & 66.9          \\
                             & \textit{mean 10}   & 70.4                                       & \textbf{75.3} & 80.8          & 69.5          \\
                             & \textit{first 120} & \textbf{72.5}                              & \textbf{75.3} & \textbf{83.6} & \textbf{70.8} \\
    \bottomrule
  \end{tabular}
\end{table}

Tables~\ref{tab:dur_mus_by_bin} and~\ref{tab:dur_ali_by_bin} quantify how
predicted scores change across inference strategies. We compare configurations
using Linear Correlation Coefficient (LCC) and Root Mean Square Error (RMSE).
We also report summary statistics of the reference scores using the standard
\textit{first 120} to contextualize absolute differences.

Across both Musicality and Alignment, increasing duration is associated with
higher RMSE and lower LCC when comparing short-context to long-context
predictions, indicating that short excerpts become less predictive of long-context
scores as track length grows. When comparing \textit{first 10} against \textit{first
  120}, the overall RMSE-to-STD ratios are 0.5575 (Musicality) and 0.6133 (Alignment),
suggesting that duration mismatch introduces substantial variation relative to
the natural score spread and motivating longer-context inference. Notably,
Alignment exhibits lower short-vs-long correlation than Musicality, consistent
with alignment requiring broader temporal context. Additionally, the correlation
for short time and long time is lower for alignment, indicating that alignment
needs longer context.

Finally, while \textit{first 10} exhibits noticeably weaker agreement with
\textit{first 120} (lower LCC and higher RMSE), it still attains reasonably
strong performance on several subsets, suggesting that short-context inference
can be ``good enough'' for some downstream uses (e.g., providing a coarse
preference signal during post-training). At the same time, the gap to \textit{first
  120} indicates that longer temporal context remains beneficial, and we leave it
to future work to develop reward models that more reliably leverage long-range
structure and text--music relationships.

\begin{table}[htbp]
  \centering
  \caption{Effect of inference duration on predicted Musicality (RMSE \& LCC between
    three configurations). Statistics for \textit{first 120} is shown }
  \label{tab:dur_mus_by_bin} \footnotesize
  \setlength{\tabcolsep}{3.5pt}
  \resizebox{\linewidth}{!}{
    \begin{tabular}{lrrrrr}
      \toprule \textbf{Pair}        & \textbf{[10,30)}    & \textbf{[30,60)}   & \textbf{[60,120)}  & \textbf{[120,500]}  & \textbf{Overall}    \\
                                    & \textit{n}\,=\,1311 & \textit{n}\,=\,824 & \textit{n}\,=\,346 & \textit{n}\,=\,1639 & \textit{n}\,=\,4120 \\
      \midrule \multicolumn{6}{l}{\textit{RMSE}}                                                                                                \\
      \textit{f10} vs \textit{f120} & 0.3022              & 0.3960             & 0.5465             & 0.7734              & 0.5687              \\
      \textit{m10} vs \textit{f120} & 0.2289              & 0.2067             & 0.4292             & 0.5014              & 0.3751              \\
      \textit{f10} vs \textit{m10}  & 0.2836              & 0.3583             & 0.3485             & 0.4478              & 0.3758              \\
      \midrule \multicolumn{6}{l}{\textit{LCC}}                                                                                                 \\
      \textit{f10} vs \textit{f120} & 0.9597              & 0.9318             & 0.8234             & 0.7541              & 0.8728              \\
      \textit{m10} vs \textit{f120} & 0.9790              & 0.9837             & 0.9292             & 0.9271              & 0.9597              \\
      \textit{f10} vs \textit{m10}  & 0.9619              & 0.9445             & 0.8811             & 0.8153              & 0.9159              \\
      \midrule \multicolumn{6}{l}{\textit{Ref (\textit{first120}) score stats}}                                                                 \\
      mean$\pm$std                  & 0.92$\pm$1.06       & 1.11$\pm$1.09      & 1.71$\pm$0.75      & 1.87$\pm$0.74       & 1.40$\pm$1.02       \\
      \textbf{[min, max]}           &                     &                    &                    &                     & [-2.31, 4.50]       \\
      \bottomrule
    \end{tabular}
  }
\end{table}

\begin{table}[htbp]
  \centering
  \caption{Effect of inference duration on predicted Text-Music Alignment (RMSE
    \& LCC between three configurations). Statistics for \textit{first 120} is
    shown }
  \label{tab:dur_ali_by_bin} \footnotesize
  \setlength{\tabcolsep}{3.5pt}
  \resizebox{\linewidth}{!}{
    \begin{tabular}{lrrrrr}
      \toprule \textbf{Pair}        & \textbf{[10,30)}    & \textbf{[30,60)}   & \textbf{[60,120)}  & \textbf{[120,500]}  & \textbf{Overall}    \\
                                    & \textit{n}\,=\,1311 & \textit{n}\,=\,824 & \textit{n}\,=\,346 & \textit{n}\,=\,1639 & \textit{n}\,=\,4120 \\
      \midrule \multicolumn{6}{l}{\textit{RMSE}}                                                                                                \\
      \textit{f10} vs \textit{f120} & 0.2726              & 0.3104             & 0.4164             & 0.6110              & 0.4539              \\
      \textit{m10} vs \textit{f120} & 0.2084              & 0.1777             & 0.3082             & 0.3410              & 0.2727              \\
      \textit{f10} vs \textit{m10}  & 0.2568              & 0.2922             & 0.2928             & 0.4022              & 0.3311              \\
      \midrule \multicolumn{6}{l}{\textit{LCC}}                                                                                                 \\
      \textit{f10} vs \textit{f120} & 0.9480              & 0.9255             & 0.8329             & 0.6836              & 0.8417              \\
      \textit{m10} vs \textit{f120} & 0.9708              & 0.9764             & 0.9309             & 0.9062              & 0.9490              \\
      \textit{f10} vs \textit{m10}  & 0.9508              & 0.9367             & 0.8828             & 0.7738              & 0.8933              \\
      \midrule \multicolumn{6}{l}{\textit{Ref (\textit{first120}) score stats}}                                                                 \\
      mean$\pm$std                  & 1.05$\pm$0.85       & 1.27$\pm$0.80      & 1.54$\pm$0.62      & 1.54$\pm$0.55       & 1.33$\pm$0.74       \\
      \textbf{[min, max]}           &                     &                    &                    &                     & [-1.76, 3.72]       \\
      \bottomrule
    \end{tabular}}
\end{table}

\section{Discussions on Musicality}

\subsection{Musicality on Music Arena}
\label{Append:Musicality} To understand how musicality and alignment jointly
dictate real-world user preferences, we regress these two predicted dimensions
against the overall human preference labels collected from Music Arena. Since
Music Arena only provides a single, holistic preference label per pair, we
evaluate whether users lean more towards aesthetic musicality or strict
instructional alignment.

\begin{table}[H]
  \centering
  \caption{Regression analysis of overall human preference on Music Arena
    using predicted Musicality (Mus) and Alignment (Ali) scores. The results indicate
    that general user preference is predominantly driven by musicality.}
  \label{tab:music_arena_regression} \resizebox{\columnwidth}{!}{%
    \begin{tabular}{lccc}
      \toprule \textbf{Method}                                  & \textbf{Accuracy (\%)} & \textbf{AUC} & \textbf{MSE} \\
      \midrule \textbf{Single Metric Threshold (Threshold = 0)} &                        &              &              \\
      Musicality Diff                                           & \textbf{73.4}          & 0.8001       & 0.1884       \\
      Alignment Diff                                            & 69.7                   & 0.7714       & 0.2031       \\
      Mus + Ali Diff (Equal)                                    & 72.6                   & 0.7962       & 0.1851       \\
      \midrule \textbf{Regression Models (5-fold CV)}           &                        &              &              \\
      Logistic Regression                                       & 73.1                   & 0.7991       & 0.1845       \\
      SVM-Linear                                                & 72.5                   & 0.7990       & 0.1846       \\
      SVM-RBF                                                   & 72.6                   & 0.7798       & 0.1875       \\
      \bottomrule
    \end{tabular}%
  }
\end{table}

Table~\ref{tab:music_arena_regression} demonstrates that musicality has an overwhelmingly
dominant influence on general preference. A simple thresholding of the
musicality difference already yields strong performance, while regression models
provide no gains (e.g., -0.3\% in accuracy) due to fold-level noise.

This dominance is directly reflected in the fitted model weights:
\begin{itemize}
  \item \textbf{Logistic Regression:} $\text{coef}_{\text{mus}}=1.2296$, $\text{coef}
          _{\text{ali}}=0.1991$, $\text{intercept}=0.0592$.

  \item \textbf{Linear SVM:} $\text{coef}_{\text{mus}}=1.2229$, $\text{coef}_{\text{ali}}
          =0.1861$.
\end{itemize}

This near-optimal alignment with in-the-wild user behavior echoes our annotation
protocols.

\subsection{Musicality: Intrinsic Audio Quality vs. Contextual Preference}
In existing datasets and models such as SongEval~\cite{songeval}, musicality
is primarily treated as an absolute measure of ``intrinsic audio quality''. These
metrics are typically \textit{reference-free}, focusing exclusively on the audio
itself. However, in the context of music generation, musicality often
interacts—or even interferes—with the generation context, specifically the multimodal
prompts fed into the model.

Compared to passive music listening, when generating music, a user's perception
of whether a piece is ``musical'' is rarely isolated. Instead, it is inherently
influenced by the aesthetic expectations established by their initial creative
intent. Therefore, we argue that evaluating musicality in a practical AIGC workflow
requires shifting from a context-free measurement to a \textit{context-aware}
judgment.

\subsubsection{Empirical Observations on Prompt Contribution}
To empirically validate whether prompt information assists in predicting human
musicality preferences, we conducted an ablation study using the CMI-Pref dataset.
We compared two variants of our model: (1) a variant that drops all prompt conditions
during both training and inference (effectively a reference-free evaluator),
and (2) a variant utilizing regular conditional training.

As shown in Table~\ref{tab:musicality_condition}, our experimental results reveal
several key findings:
\begin{itemize}
  \item \textbf{Performance Gain from Context:} Incorporating prompt
        conditions significantly improves the overall musicality prediction
        accuracy from 70.20\% to 75.60\% ($\Delta = +5.40\%$). Specifically, the
        text-only setting also shows a marginal gain (+1.60\%).

  \item \textbf{Reference Audio as a Strong Anchor:} The improvement is most
        pronounced when reference audio is present. The ``Text + Audio + Lyrics'' and
        ``Text + Audio w/o Lyrics'' modalities yield substantial gains of +13.60\%
        and +10.40\%, respectively. This suggests that reference-based context provides
        a much clearer signal for aesthetic preference than text alone.

  \item \textbf{Modality Bottleneck in Lyrics:} Conversely, the ``Text + Lyrics
        w/o Audio'' setting is the only modality that shows a decline (-4.00\%).
        We attribute this to the ineffectiveness of the frozen MuQ-MuLan text
        encoder, which struggles to process raw lyric structures without acoustic grounding,
        thereby introducing semantic noise rather than useful context.

  \item \textbf{Comparison with Reference-Free Baselines:} As shown in our
        main results, Table \ref{tab:musicality-all-data-statis}, reference-free
        metrics like SongEval-RM show competitive performance on general acoustic
        quality but are outperformed by our context-aware model on CMI-Pref. This highlights
        the limitation of condition-free evaluation in complex compositional tasks.
\end{itemize}

\begin{table}[htbp]
  \centering
  \caption{Ablation study on prompt conditions for Musicality prediction on
    CMI-Pref. We compare models trained and evaluated with and without compositional
    conditions across different prompt modalities ($n=125$ for each subset).}
  \label{tab:musicality_condition} \resizebox{\columnwidth}{!}{%
    \begin{tabular}{lccc}
      \toprule \textbf{Prompt Modality}   & \textbf{w/ Condition Acc (\%)} & \textbf{w/o Condition Acc (\%)} & $\Delta$ \\
      \midrule Text + Audio + Lyrics      & \textbf{82.40}                 & 68.80                           & +13.60   \\
      Text + Audio w/o Lyrics             & \textbf{79.20}                 & 68.80                           & +10.40   \\
      Text + Lyrics w/o Audio             & 72.00                          & \textbf{76.00}                  & -4.00    \\
      Text only                           & \textbf{68.80}                 & 67.20                           & +1.60    \\
      \midrule \textbf{Overall ($N=500$)} & \textbf{75.60}                 & 70.20                           & +5.40    \\
      \bottomrule
    \end{tabular}%
  }
\end{table}

\subsubsection{Hypotheses: Why Does Context Assist Musicality Prediction?}
We propose two primary hypotheses to explain why the inclusion of
prompts—which are in theory independent of absolute audio quality—enhances the
model's predictive performance:

\textbf{1. Inherent Association in Human Judgment:} During annotation, human
experts may find it difficult to completely decouple musicality from alignment.
For instance, a prompt specifying a ``lo-fi aesthetic'' might lead an
annotator to perceive low-fidelity audio as a musical choice rather than a
technical flaw. Thus, the prompt acts as a ``taste anchor'' that recalibrates
the aesthetic scale.

\textbf{2. Inference Assistance via Non-trivial Shortcuts:} From the data correlation
perspective, there is an undeniable, intrinsic correlation between the prompt and
the evaluated audio, as the latter is directly generated conditioned on the
former from a music generation model. Because of this inherent link, the
prompt provides critical contextual clues about the expected acoustic features.
By attending to the prompt, the model can anticipate the intended musical
elements and more easily identify task-specific artifacts that degrade quality
but are difficult to detect from the raw audio alone.

\subsubsection{Implications for Evaluating Music Creation}
Regardless of whether these gains stem from human psychological anchoring or
model-side statistical correlation, the results highlight a crucial distinction
between \textit{music listening} and \textit{music creation}. While standalone
acoustic metrics are sufficient for evaluating isolated tracks, the evaluation
of AIGC music creation fundamentally requires context. Therefore, \textbf{multimodal
  inputs (Text, Lyrics, Audio) should be treated as an integral part of the
  evaluator's state}. For reward models in these tasks, transitioning from
absolute acoustic measurement to context-aware preference is essential for
achieving closer alignment with real-world user intent.

\section{Statistical Details}
\label{app:statistical-details}

\subsection{Significance of Benchmark Improvements}

We report additional significance checks from the rebuttal for key
preference-based gains. These tests compare our model with strong baselines on
\textsc{CMI-Pref} and Music Arena. The following checks indicate that the main
improvements are unlikely to be explained by chance:

\begin{enumerate}[leftmargin=*, itemsep=0.2em]
  \item \textbf{CMI-Pref musicality}: CMI-Pref vs Gemini2.5-Pro, 78.60\% vs
        70.00\%, $p=9.30\times10^{-4}$.
  \item \textbf{CMI-Pref musicality}: CMI-Pref vs audiobox-PQ, 78.60\% vs
        73.80\%, $p=3.74\times10^{-2}$.
  \item \textbf{Music Arena musicality}: CMI+MusicEval vs Gemini2.5-Pro,
        73.21\% vs 69.75\%, $p=2.37\times10^{-2}$.
  \item \textbf{High-confidence musicality split}: CMI-Pref vs Gemini2.5-Pro,
        85.71\% vs 75.96\%, $p=1.09\times10^{-3}$.
\end{enumerate}

\subsection{Human Listening Significance for Test-Time Scaling}
\label{app:stat_testtime}
For test-time scaling (Sec.~4.3), objective reranking uses 2,800 prompts from
MusicGen-small and Stable-Audio-Open-small. Subjective listening uses
\textbf{50 prompts} and \textbf{four ranks} (GT, Top-10, Top-3, Top-1), yielding
200 generations and 300 pairwise comparisons. Following the rebuttal protocol,
ties are split equally and significance is computed with an exact one-sided
binomial test on effective non-tied comparisons.

Table~\ref{tab:testtime_pvalues} shows that the largest ranking gaps are
significant, while differences among reranked candidates are often not
significant. This matches Fig.~\ref{fig:heatmaps}, where GT is consistently
preferred and Top-3/Top-10 are much closer.

\begin{table}[t]
  \centering
  \small
  \caption{Significance checks for human listening results in test-time
    scaling.}
  \label{tab:testtime_pvalues}
  \resizebox{\columnwidth}{!}{
    \begin{tabular}{lll}
      \toprule
      \textbf{Backbone / Comparison} & \textbf{Ordering} & \textbf{$p$-value} \\
      \midrule
      Stable Audio                   & GT $>$ Top-10     & 0.003              \\
      MusicGen                       & GT $>$ Top-1      & 0.0001             \\
      Stable Audio                   & GT $>$ Top-1      & 0.0001             \\
      MusicGen                       & Top-10 $>$ Top-3  & 0.016              \\
      MusicGen                       & Top-3 $>$ Top-1   & 0.101              \\
      Stable Audio                   & Top-10 $>$ Top-3  & 0.890              \\
      Stable Audio                   & Top-3 $>$ Top-1   & 0.239              \\
      \bottomrule
    \end{tabular}
  }
\end{table}

\section{Detailed Analysis of Results}
\label{app:result}

\subsection{Lyrics Transcription as a Proxy Metric}
\label{app:lyrics-wer}

We further test whether explicit lyric transcription can serve as a simple proxy
for lyric-following preference. On the 250 lyric-conditioned pairs in CMI-Pref-test,
we transcribe all 500 audio samples using Whisper ASR and compute word error
rate (WER) against the prompt lyrics. A WER-based preference rule selects the
candidate with lower WER; when both candidates obtain identical WER, the comparison
is treated as a tie and excluded from WER-based decision accuracy, yielding 225
non-tie comparisons.

WER-based preference prediction reaches only 60.0\% accuracy, substantially below
the 76.0\% accuracy of CMI-RM on the same lyric-conditioned setting. This result
suggests that transcription quality is a useful auxiliary signal, but not a sufficient
predictor of human preference for lyric-conditioned music. Human annotators also
consider vocal naturalness, prosody, musicality, structural coherence, and how
well the lyrics are integrated with the requested musical style, which are not
captured by WER alone.

\subsection{Performance on Music Arena Subsets}
\begin{table*}
  [ht]
  \caption{Benchmark Musicality ACC Results on Music Arena (Time \& Data Type Analysis).}
  \label{tab:cmu-arena-simplified}
  \centering
  \scriptsize
  \setlength{\tabcolsep}{6pt} %
  \renewcommand{\arraystretch}{1.2}

  \begin{tabular}{l|c|ccccc|cc}
    \toprule \multirow{2}{*}{\textbf{Model}} & \textbf{Total}    & \multicolumn{5}{c|}{\textbf{Time Period (Month)}} & \multicolumn{2}{c}{\textbf{Data Type}}                                                                                                     \\
                                             & (\%)              & Jul-Aug                                           & Sep                                    & Oct               & Nov               & Dec               & Instru            & Vocal             \\
    \midrule

    PAM score                                & 63.13             & 69.13                                             & 75.41                                  & 58.56             & 53.14             & 55.27             & 56.84             & 68.35             \\
    audiobox-CE                              & 64.25             & 67.79                                             & 73.77                                  & 63.36             & 60.14             & 55.27             & 64.25             & 64.25             \\
    audiobox-CU                              & 67.76             & 71.36                                             & \underline{77.0}                       & 66.78             & 60.14             & 60.73             & 68.04             & 67.53             \\
    audiobox-PC                              & 58.73             & 63.75                                             & 70.49                                  & 55.82             & 54.54             & 48.00             & 52.88             & 63.57             \\
    audiobox-PQ                              & 67.54             & 68.01                                             & 74.86                                  & 67.47             & 65.04             & 63.27             & 68.04             & 67.12             \\
    SongEval-RM                              & \textbf{73.88}    & \underline{73.60}                                 & \textbf{78.69}                         & 71.92             & 64.34             & \textbf{ 78.18}   & \textbf{77.27}    & \underline{71.08} \\
    Omni-Reward                              & 54.02             & 49.44                                             & 49.18                                  & 60.61             & 53.85             & 57.81             & 54.53             & 53.62             \\
    \midrule Qwen2-audio                     & 35.99             & 38.26                                             & 37.91                                  & 31.23             & 38.24             & 34.91             & 30.66             & 39.63             \\
    Qwen2.5-Omni                             & 36.05             & 40.04                                             & 39.89                                  & 38.36             & 24.48             & 30.55             & 29.98             & 41.23             \\
    Qwen3-Omni                               & 59.63             & 59.06                                             & 58.47                                  & 59.93             & 60.84             & 60.36             & 62.93             & 56.89             \\
    Gemini2.5-Flash                          & 64.12             & 60.54                                             & 62.84                                  & 69.52             & 61.27             & 66.55             & 65.84             & 62.70             \\
    Gemini2.5-Pro                            & 69.75             & 65.99                                             & 64.48                                  & 71.88             & \underline{73.24} & \underline{75.27} & \underline{75.78} & 64.69             \\
    Gemini3-Pro                              & 68.85             & 63.59                                             & 60.28                                  & \underline{74.31} & \textbf{78.63}    & 72.22             & 76.03             & 62.46             \\
    \midrule - w/o f.t.: Distill only        & 65.37             & 67.56                                             & 68.3                                   & 67.8              & 68.53             & 55.64             & 63.26             & 67.12             \\
    \makecell[l]{- w/ f.t.: CMI-Pref}        & {71.57}           & 71.36                                             & 75.41                                  & 72.95             & 65.04             & 71.27             & 73.81             & 69.71             \\
    \makecell[l]{- w/ f.t.: CMI + MusicEval} & \underline{73.21} & \textbf{74.27}                                    & 75.96                                  & \textbf{74.66}    & 65.73             & 72.00             & 74.14             & \textbf{72.44}    \\
    \bottomrule
  \end{tabular}%
  \vskip -0.1in
\end{table*}

Tables \ref{tab:cmu-arena-simplified} and \ref{tab:ablation-results} reveal two
dominant factors that influence reward model performance: temporal
distribution shift and annotation confidence.

Table \ref{tab:cmu-arena-simplified} shows a clear temporal drift in Music
Arena. Several reference-free metrics, particularly the Audiobox variants, exhibit
a noticeable decline in accuracy from early to later months, indicating
sensitivity to evolving generation quality and user preference distributions. SongEval-RM
demonstrates relatively stronger stability across time, while general-purpose
multimodal LLMs show substantial month-to-month variance, suggesting limited robustness
for fine-grained musicality judgment. In contrast, our models maintain more consistent
performance over time and achieve improved robustness on vocal music,
especially when jointly fine-tuned with MusicEval, highlighting the benefit of
music-specific supervision.
\subsection{Performance on CMI-Pref Subsets}
Table \ref{tab:ablation-results} analyzes performance on the CMI-Pref test set
by stratifying samples according to annotator-reported confidence, which in our
data collection protocol explicitly measures the perceived \emph{preference
  margin} between two candidates. Unlike traditional pairwise annotation schemes
that only record a binary choice, annotators are required to select a
preferred sample and additionally indicate how strongly one sample is
preferred over the other.

\begin{table*}
  [ht]
  \caption{Benchmark Musicality ACC Results on CMI-Pref (Ablation Analysis).}
  \label{tab:ablation-results}
  \centering
  \scriptsize

  \begin{tabular}{l|c|ccc|cc}
    \toprule \multirow{2}{*}{Musicality}     & Total   & \multicolumn{3}{c|}{Confidence Level} & \multicolumn{2}{c}{Data Type}                                                    \\
                                             & (500)   & Conf $<3$ (66)                        & Conf $=3$ (128)               & Conf $>3$ (306) & Instru (250)     & Vocal (250) \\
    \midrule

    PAM score                                & 65.40\% & 63.64\%                               & 64.84\%                       & 66.01\%         & 67.60\%          & 63.20\%     \\
    audiobox-CE                              & 71.80\% & 63.64\%                               & 72.66\%                       & 73.20\%         & 70.80\%          & 72.80\%     \\
    audiobox-CU                              & 71.40\% & \underline{66.67\%}                   & 71.88\%                       & 72.22\%         & 68.80\%          & 74.00\%     \\
    audiobox-PC                              & 59.00\% & 46.97\%                               & 54.69\%                       & 63.40\%         & 60.00\%          & 58.00\%     \\
    audiobox-PQ                              & 73.80\% & \textbf{74.24\%}                      & \textbf{78.13\%}              & 71.90\%         & 73.20\%          & 74.40\%     \\
    SongEval-RM                              & 72.40\% & 63.64\%                               & 71.09\%                       & 74.84\%         & 70.80\%          & 74.00\%     \\
    Omni-Reward                              & 65.60\% & 56.06\%                               & 66.41\%                       & 67.32\%         & 59.20\%          & 72.00\%     \\
    \midrule Qwen2-audio                     & 8.60\%  & 4.54\%                                & 7.81\%                        & 9.80\%          & 5.20\%           & 12.00\%     \\
    Qwen2.5-omni                             & 17.40\% & 15.15\%                               & 16.41\%                       & 18.40\%         & 15.20\%          & 19.60\%     \\
    Qwen3-omni                               & 60.40\% & 54.55\%                               & 53.12\%                       & 64.70\%         & 63.20\%          & 57.60\%     \\
    Gemini2.5-flash                          & 64.20\% & 56.57\%                               & 57.94\%                       & 71.94\%         & 74.40\%          & 54.00\%     \\
    Gemini2.5-pro                            & 70.00\% & 48.49\%                               & 69.29\%                       & 75.96\%         & \textbf{78.80\%} & 61.20\%     \\
    Gemini3-pro                              & 65.80\% & 57.94\%                               & 66.67\%                       & 72.67\%         & 73.20\%          & 58.40\%     \\
    \midrule - w/o f.t.: Distill only        & 70.80\% & 53.03\%                               & 67.19\%                       & 76.14\%         & 65.60\%          & 76.00\%     \\ %
    \makecell[l]{- w/ f.t.: CMI-Pref}        & 77.80\% & 66.67\%                               & 76.56\%                       & 80.72\%         & 74.00\%          & 81.60\%     \\ %
    \makecell[l]{- w/ f.t.: CMI + MusicEval} & 78.10\% & 66.67\%                               & 75.39\%                       & 81.70\%         & 75.60\%          & 80.60\%     \\ %
    \bottomrule
  \end{tabular}%
  \vskip -0.1in
\end{table*}

We observe a clear monotonic relationship between confidence and model accuracy
across all methods. Higher-confidence comparisons, corresponding to larger perceptual
gaps in musical quality or instruction adherence, are consistently easier to
predict. Lower-confidence comparisons reflect fine-grained distinctions with smaller
margins, which naturally impose a more challenging learning problem.

Importantly, our proposed reward models exhibit the largest performance gains in
the high-confidence regime. The CMI-Pref fine-tuned model substantially
outperforms all baselines when confidence is high, indicating strong alignment
with clear and decisive human preferences. At low confidence levels, performance
differences across methods narrow, suggesting that improvements on near-tie
comparisons are fundamentally limited by the small preference margins rather
than model capacity.

\begin{table*}
  [ht]
  \caption{Benchmark Text-Music Alignment ACC Results on CMI-Pref(Ablation
    Analysis).}
  \label{tab:text-music-acc}
  \centering
  \scriptsize
  \begin{tabular}{l|c|ccc|cc}
    \toprule \multirow{2}{*}{Text-Music Alignment} & Total   & \multicolumn{3}{c|}{Confidence Level} & \multicolumn{2}{c}{Data Type}                                                \\
                                                   & (250)   & Conf $<3$ (47)                        & Conf $=3$ (58)                & Conf $>3$ (145) & Instru (125) & Vocal (125) \\
    \midrule

    audiobox-CE                                    & 60.00\% & 45.95\%                               & 63.64\%                       & 61.90\%         & 56.80\%      & 63.20\%     \\
    audiobox-CU                                    & 59.60\% & 54.05\%                               & 63.64\%                       & 59.18\%         & 53.60\%      & 65.60\%     \\
    audiobox-PC                                    & 56.00\% & 35.13\%                               & 59.09\%                       & 59.86\%         & 55.20\%      & 56.80\%     \\
    audiobox-PQ                                    & 59.60\% & 54.05\%                               & 65.15\%                       & 58.50\%         & 55.20\%      & 64.00\%     \\
    CLAP score                                     & 62.40\% & 51.35\%                               & 59.09\%                       & 66.67\%         & 60.80\%      & 64.00\%     \\
    CLAP music score                               & 70.40\% & 75.68\%                               & 60.61\%                       & 73.47\%         & 67.20\%      & 73.60\%     \\
    MuQ-Mulan                                      & 66.40\% & 67.56\%                               & 60.00\%                       & 69.39\%         & 64.80\%      & 68.00\%     \\
    CLAMP3 score                                   & 62.80\% & 64.86\%                               & 60.61\%                       & 63.27\%         & 63.20\%      & 62.40\%     \\
    Omni-Reward                                    & 68.80\% & 57.14\%                               & 65.96\%                       & 72.67\%         & 67.20\%      & 70.40\%     \\
    \midrule Qwen2-audio                           & 1.60\%  & 2.70\%                                & 1.51\%                        & 1.36\%          & 0.80\%       & 2.40\%      \\
    Qwen2.5-Omni                                   & 34.40\% & 27.03\%                               & 33.33\%                       & 36.73\%         & 31.20\%      & 37.60\%     \\
    Qwen3-Omni                                     & 63.60\% & 56.76\%                               & 63.64\%                       & 66.67\%         & 67.20\%      & 60.00\%     \\
    Gemini2.5-Flash                                & 60.80\% & 48.65\%                               & 60.61\%                       & 63.95\%         & 65.60\%      & 56.00\%     \\
    Gemini2.5-Pro                                  & 67.20\% & 54.05\%                               & 66.67\%                       & 70.75\%         & 71.20\%      & 63.20\%     \\
    Gemini3-Pro                                    & 64.00\% & 54.05\%                               & 66.67\%                       & 65.31\%         & 67.20\%      & 60.80\%     \\
    \midrule - w/o f.t.: Distill only              & 69.60\% & 65.96\%                               & 70.69\%                       & 70.34\%         & 67.20\%      & 72.00\%     \\ %
    \makecell[l]{- w/ f.t.: CMI-Pref}              & 68.80\% & 55.32\%                               & 74.14\%                       & 71.03\%         & 64.80\%      & 72.80\%     \\ %
    \makecell[l]{- w/ f.t.: CMI + MusicEval}       & 70.20\% & 51.06\%                               & 68.96\%                       & 76.89\%         & 67.60\%      & 72.80\%     \\
    \bottomrule
  \end{tabular}%
  \vskip -0.1in
  \vspace{15pt}
  \caption{Benchmark Audio-Music Alignment ACC Results on CMI-Pref (Ablation Analysis).}
  \label{tab:audio-music-acc} \scriptsize
  \begin{tabular}{l|c|ccc|cc}
    \toprule \multirow{2}{*}{Audio-Music Alignment} & Total   & \multicolumn{3}{c|}{Confidence Level} & \multicolumn{2}{c}{Data Type}                                                \\
                                                    & (250)   & Conf $<3$ (42)                        & Conf $=3$ (47)                & Conf $>3$ (161) & Instru (125) & Vocal (125) \\
    \midrule
    OmniRewardModel                                 & 68.80\% & 57.14\%                               & 65.96\%                       & 72.67\%         & 67.20\%      & 70.40\%     \\
    Qwen2-audio                                     & 11.60\% & 11.90\%                               & 4.84\%                        & 14.38\%         & 8.80\%       & 14.40\%     \\
    Qwen2.5-Omni                                    & 28.80\% & 26.19\%                               & 25.81\%                       & 30.82\%         & 25.60\%      & 32.00\%     \\
    Qwen3-Omni                                      & 64.00\% & 54.76\%                               & 61.29\%                       & 67.81\%         & 64.80\%      & 63.20\%     \\
    Gemini2.5-Flash                                 & 62.00\% & 50.00\%                               & 61.29\%                       & 65.75\%         & 69.60\%      & 54.40\%     \\
    Gemini2.5-Pro                                   & 72.80\% & 52.38\%                               & 72.58\%                       & 78.77\%         & 73.60\%      & 72.00\%     \\
    Gemini3-Pro                                     & 66.80\% & 61.90\%                               & 53.23\%                       & 73.97\%         & 68.80\%      & 64.80\%     \\
    \midrule - w/o f.t.: Distill only               & 73.20\% & 61.90\%                               & 63.83\%                       & 78.88\%         & 69.20\%      & 77.20\%     \\
    \makecell[l]{- w/ f.t.: CMI-Pref}               & 79.20\% & 66.67\%                               & 72.34\%                       & 84.47\%         & 76.00\%      & 82.40\%     \\ %
    \makecell[l]{- w/ f.t.: CMI + MusicEval}        & 77.80\% & 55.95\%                               & 78.72\%                       & 83.22\%         & 76.40\%      & 79.20\%     \\
    \bottomrule
  \end{tabular}%
  \vskip -0.1in
\end{table*}

Overall, these results demonstrate that explicitly modeling preference
strength during data collection provides a principled way to analyze reward
model behavior beyond binary accuracy, and highlights the effectiveness of our
approach in capturing dominant human preference signals relevant for downstream
reranking and selection.

\clearpage

\section{Leaderboard for music generation models}
We report an overall leaderboard of music generation models evaluated by our
trained reward model, \textsc{CMI-RM}. All audio generations come from \textsc{CMI-Pref-Psuedo},
where each model is queried with the same pool of prompts (and, when applicable,
the same audio prompts). For every generated sample, we run \textsc{CMI-RM}
with the prompt (text instruction and optional audio prompt) and the generated
audio as inputs, and obtain two scalar signals: \textit{alignment} and \textit{musicality}.

To aggregate results into a unified ranking, we construct prompt-wise full round-robin
comparisons: for each unique prompt (under a fixed input type), we enumerate all
pairs of model outputs generated from that same prompt. For each input type (Inst./Song
$\times$ w/ audio / w/o audio) and each dimension (alignment / musicality), \textsc{CMI-RM}
assigns a binary label for each pair (i.e., whether model $a$ is preferred over
model $b$ under the same prompt). We then compute a Bradley--Terry (BT) style
ranking (equivalently, an Elo-like rating variant) by aggregating these pairwise
win/loss labels, and report the resulting rank scores in
\autoref{tab:eval_ranks_sorted}. The corresponding ranking scores are reported
in \autoref{tab:eval_scores_sorted}. Inst.\ denotes instrumental music generation,
Song denotes song generation, and ``w/ audio'' indicates that an audio prompt
is provided in addition to the text instruction; modalities not supported by a
model are left blank. For readability, we linearly rescale scores as $\text{score}
  \times 400 + 1500$. Proprietary models (no public checkpoints) are marked in
\textit{italics}. Note that in Song (w/ audio), \textit{mureka-o2} and \textit{minimax-music-2}
have fewer generations (see \autoref{tab:gencount}); hence their apparent advantage
in this subset may be less reliable.

We observe that:
\begin{enumerate}
  \item The gap between open-weight and closed-source models remains large:
        the top-5 models in each metric are dominated by proprietary systems.

  \item In pure song generation, the gap between Suno and other proprietary
        models is narrowing. In modalities with abundant generations, Minimax
        Music and Mureka outperform several Suno variants.

  \item Recent open-source models---\textsc{Levo} for song generation, \textsc{Magenta
          Realtime} for instrumental generation, and \textsc{AceStep} across both---exhibit
        competitive performance.
\end{enumerate}

\begin{table}[t]
  \centering
  \caption{Generation counts per model by modality }
  \label{tab:gencount} \resizebox{\linewidth}{!}{
    \begin{tabular}{lcccc}
      \toprule                                   & \multicolumn{2}{c}{Instrumental} & \multicolumn{2}{c}{Song}                        \\
      \cmidrule(lr){2-3}\cmidrule(lr){4-5} Model & w/ audio                         & w/o audio                & w/ audio & w/o audio \\
      \midrule \textit{suno-v5}                  & 212                              & 221                      & 213      & 288       \\
      \textit{mureka-o2}                         & 0                                & 0                        & 15       & 163       \\
      \textit{suno-v4.5-plus}                    & 214                              & 221                      & 216      & 288       \\
      \textit{mureka-v7.5}                       & 7                                & 117                      & 8        & 211       \\
      \textit{minimax-music-2}                   & 0                                & 0                        & 27       & 429       \\
      \textit{suno-v4.5}                         & 165                              & 193                      & 163      & 260       \\
      \textit{suno-v3.5}                         & 0                                & 399                      & 0        & 378       \\
      \textit{suno-v4}                           & 193                              & 193                      & 184      & 260       \\
      levo                                       & 0                                & 0                        & 484      & 735       \\
      magenta-rt-large                           & 2979                             & 3037                     & 0        & 0         \\
      \textit{satwo}                             & 445                              & 320                      & 0        & 0         \\
      sao                                        & 1348                             & 4674                     & 0        & 0         \\
      \textit{loudly-music}                      & 10                               & 248                      & 0        & 0         \\
      acestep                                    & 1346                             & 4675                     & 1470     & 1468      \\
      audioldm                                   & 829                              & 937                      & 0        & 0         \\
      diffrhythm                                 & 831                              & 513                      & 843      & 730       \\
      musicldm                                   & 1347                             & 4678                     & 0        & 0         \\
      jamify                                     & 580                              & 341                      & 496      & 455       \\
      yue                                        & 0                                & 0                        & 855      & 898       \\
      sao-small                                  & 830                              & 938                      & 0        & 0         \\
      musicgen-medium                            & 830                              & 937                      & 0        & 0         \\
      audioldm2-music                            & 1348                             & 4677                     & 0        & 0         \\
      songgen                                    & 0                                & 0                        & 2125     & 2130      \\
      \bottomrule
    \end{tabular}}%
\end{table}

\begin{table*}
  [t]
  \centering
  \caption{Per-column ranks (1=best). Inst. refers to instrumental music
    generation, Song referes to song generation, w/audio indicates if an audio
    prompt is used. Unavailable modalities are left out.}
  \label{tab:eval_ranks_sorted} \resizebox{\textwidth}{!}{%
    \begin{tabular}{lcccccccc}
      \toprule                                                                       & \multicolumn{2}{c}{Inst.\ (w/ audio)} & \multicolumn{2}{c}{Inst.\ (w/o audio)} & \multicolumn{2}{c}{Song\ (w/ audio)} & \multicolumn{2}{c}{Song\ (w/o audio)}                                                     \\
      \cmidrule(lr){2-3}\cmidrule(lr){4-5}\cmidrule(lr){6-7}\cmidrule(lr){8-9} Model & Alignment                             & Musicality                             & Alignment                            & Musicality                            & Alignment  & Musicality & Alignment  & Musicality \\
      \midrule \textit{suno-v5}                                                      & 1                                     & 1                                      & 3                                    & 3                                     & 3          & 3          & 2          & 3          \\
      \textit{mureka-o2}                                                             & -                                     & -                                      & -                                    & -                                     & 2          & 1          & 4          & 1          \\
      \textit{suno-v4.5-plus}                                                        & 2                                     & 2                                      & 1                                    & 2                                     & 4          & 4          & 3          & 5          \\
      \textit{mureka-v7.5}                                                           & 6                                     & 4                                      & 2                                    & 1                                     & 7          & 6          & 5          & 2          \\
      \textit{minimax-music-2.0}                                                     & -                                     & -                                      & -                                    & -                                     & 1          & 2          & 1          & 4          \\
      \textit{suno-v4.5}                                                             & 3                                     & 3                                      & 4                                    & 4                                     & 5          & 5          & 8          & 7          \\
      \textit{suno-v3.5}                                                             & -                                     & -                                      & 5                                    & 5                                     & -          & -          & 6          & 8          \\
      \textit{suno-v4}                                                               & 7                                     & 5                                      & 6                                    & 6                                     & 8          & 8          & 7          & 6          \\
      levo                                                                           & -                                     & -                                      & -                                    & -                                     & \textbf{6} & \textbf{7} & \textbf{9} & \textbf{9} \\
      magenta-rt-large                                                               & \textbf{5}                            & \textbf{7}                             & \textbf{7}                           & \textbf{8}                            & -          & -          & -          & -          \\
      \textit{satwo}                                                                 & 4                                     & 6                                      & 11                                   & 12                                    & -          & -          & -          & -          \\
      sao                                                                            & 8                                     & 9                                      & 8                                    & 10                                    & -          & -          & -          & -          \\
      \textit{loudly-music}                                                          & 12                                    & 10                                     & 9                                    & 7                                     & -          & -          & -          & -          \\
      acestep                                                                        & 11                                    & 11                                     & 13                                   & 11                                    & 9          & 9          & 10         & 11         \\
      audioldm                                                                       & 9                                     & 13                                     & 12                                   & 16                                    & -          & -          & -          & -          \\
      diffrhythm                                                                     & 13                                    & 12                                     & 10                                   & 9                                     & 10         & 10         & 12         & 10         \\
      musicldm                                                                       & 14                                    & 16                                     & 14                                   & 14                                    & -          & -          & -          & -          \\
      jamify                                                                         & 10                                    & 8                                      & 17                                   & 13                                    & 12         & 11         & 13         & 13         \\
      yue                                                                            & -                                     & -                                      & -                                    & -                                     & 11         & 12         & 11         & 12         \\
      sao-small                                                                      & 16                                    & 15                                     & 16                                   & 17                                    & -          & -          & -          & -          \\
      musicgen-medium                                                                & 15                                    & 14                                     & 18                                   & 18                                    & -          & -          & -          & -          \\
      audioldm2-music                                                                & 17                                    & 17                                     & 15                                   & 15                                    & -          & -          & -          & -          \\
      songgen                                                                        & -                                     & -                                      & -                                    & -                                     & 13         & 13         & 14         & 14         \\
      \bottomrule
    \end{tabular}%
  }

  \vspace{6pt}
  \caption{Scores per column computed by Bradley-Terry ranking model. Scores are
    scaled by 400 and starts at 1500. Inst. refers to instrumental music
    generation, Song referes to song generation, w/audio indicates if an audio
    prompt is used. Unavailable modalities are left out.}
  \label{tab:eval_scores_sorted}
  \resizebox{\textwidth}{!}{%
    \begin{tabular}{lcccccccc}
      \toprule                                                                       & \multicolumn{2}{c}{Inst.\ (w/ audio)} & \multicolumn{2}{c}{Inst.\ (w/o audio)} & \multicolumn{2}{c}{Song\ (w/ audio)} & \multicolumn{2}{c}{Song\ (w/o audio)}                                                                \\
      \cmidrule(lr){2-3}\cmidrule(lr){4-5}\cmidrule(lr){6-7}\cmidrule(lr){8-9} Model & Alignment                             & Musicality                             & Alignment                            & Musicality                            & Alignment        & Musicality       & Alignment & Musicality \\
      \midrule \textit{suno-v5}                                                      & 1679.68                               & 1722.06                                & 1678.19                              & 1736.51                               & 1632.66          & 1629.18          & 1646.61   & 1640.35    \\
      \textit{mureka-o2}                                                             & -                                     & -                                      & -                                    & -                                     & 1635.26          & 1724.59          & 1617.19   & 1697.13    \\
      \textit{suno-v4.5-plus}                                                        & 1655.11                               & 1695.81                                & 1712.03                              & 1749.49                               & 1616.90          & 1613.08          & 1624.70   & 1625.19    \\
      \textit{mureka-v7.5}                                                           & 1557.91                               & 1646.86                                & 1711.13                              & 1809.69                               & 1585.47          & 1589.45          & 1602.01   & 1672.76    \\
      \textit{minimax-music-2.0}                                                     & -                                     & -                                      & -                                    & -                                     & 1640.14          & 1632.54          & 1665.48   & 1634.73    \\
      \textit{suno-v4.5}                                                             & 1623.63                               & 1672.01                                & 1635.68                              & 1682.98                               & 1610.11          & 1610.45          & 1578.10   & 1601.62    \\
      \textit{suno-v3.5}                                                             & -                                     & -                                      & 1600.23                              & 1682.56                               & -                & -                & 1583.51   & 1595.66    \\
      \textit{suno-v4}                                                               & 1542.72                               & 1593.45                                & 1597.78                              & 1681.80                               & 1526.98          & 1514.34          & 1580.83   & 1617.20    \\
      levo                                                                           & -                                     & -                                      & -                                    & -                                     & \textbf{1590.79} & \textbf{1576.12} & 1518.99   & 1504.16    \\
      magenta-rt-large                                                               & 1577.06                               & 1547.47                                & 1535.96                              & 1472.59                               & -                & -                & -         & -          \\
      \textit{satwo}                                                                 & 1597.36                               & 1572.05                                & 1438.93                              & 1405.22                               & -                & -                & -         & -          \\
      sao                                                                            & 1491.48                               & 1475.49                                & 1485.93                              & 1439.23                               & -                & -                & -         & -          \\
      \textit{loudly-music}                                                          & 1444.07                               & 1472.19                                & 1461.57                              & 1475.87                               & -                & -                & -         & -          \\
      acestep                                                                        & 1445.90                               & 1446.98                                & 1426.37                              & 1427.27                               & 1453.36          & 1429.65          & 1430.32   & 1383.79    \\
      audioldm                                                                       & 1480.04                               & 1420.77                                & 1429.10                              & 1311.51                               & -                & -                & -         & -          \\
      diffrhythm                                                                     & 1442.14                               & 1428.54                                & 1457.82                              & 1456.67                               & 1363.93          & 1364.21          & 1379.93   & 1386.05    \\
      musicldm                                                                       & 1389.73                               & 1331.83                                & 1404.59                              & 1356.01                               & -                & -                & -         & -          \\
      jamify                                                                         & 1471.97                               & 1482.84                                & 1350.61                              & 1395.26                               & 1351.20          & 1355.35          & 1216.11   & 1235.35    \\
      yue                                                                            & -                                     & -                                      & -                                    & -                                     & 1354.09          & 1324.13          & 1392.80   & 1334.58    \\
      sao-small                                                                      & 1385.25                               & 1361.97                                & 1351.36                              & 1305.54                               & -                & -                & -         & -          \\
      musicgen-medium                                                                & 1388.95                               & 1362.58                                & 1329.69                              & 1271.37                               & -                & -                & -         & -          \\
      audioldm2-music                                                                & 1326.99                               & 1267.09                                & 1393.03                              & 1340.44                               & -                & -                & -         & -          \\
      songgen                                                                        & -                                     & -                                      & -                                    & -                                     & 1139.10          & 1136.92          & 1163.41   & 1071.42    \\
      \bottomrule
    \end{tabular}%
  }
\end{table*}

\clearpage

\section{LLM-as-a-Judge Prompt Architecture}
\label{app:prompts}

To leverage frontier multimodal Large Language Models (MLLMs) as evaluators
for generated music, we designed a structured message pipeline that seamlessly
integrates text descriptions, optional lyrics, optional reference audio, and
the generated audio candidates. This ensures the MLLM assesses the candidates based
on comprehensive compositional multimodal instructions (CMI) and outputs a rigorously
structured JSON response.

\subsection{Message Pipeline Construction}
The input to the MLLM API is constructed sequentially as a list of messages.
This pipeline dynamically adapts to the presence of optional modalities (lyrics,
reference audio) and experimental settings (few-shot prompting, chain-of-thought
reasoning). The exact message sequence is defined as follows:

\begin{enumerate}
  \item \textbf{Task Instruction (System)} \\
        Initializes the evaluator role, defines the evaluation criteria (musicality,
        instruction following), and specifies the expected JSON schema.
        \begin{itemize}
          \item \textbf{Role:}system

          \item \textbf{Content:} [See Section~\ref{sec:system_prompts} for specific
                  System Prompts]
        \end{itemize}

  \item \textbf{Few-Shot Examples (System, Optional)} \\
        If few-shot evaluation is enabled, an instruction and a set of predefined examples
        are appended to the message sequence. However, we disable this option during
        distillation, since they did not yield better consistency with human
        preference labels in our tests and led to a substantial increase in
        context length.
        \begin{itemize}
          \item \textbf{Role:}system

          \item \textbf{Content:} ``You will now be shown several examples of
                audio comparisons to help you understand how to evaluate the audios.''
                \textit{(Followed by few-shot message pairs)}
        \end{itemize}

  \item \textbf{User Instruction Context (User)} \\
        Injects the specific CMI used for generation.
        \begin{itemize}
          \item \textbf{Role:}user

          \item \textbf{Content:}
                \begin{itemize}
                  \item $[$Text$]$ ``The user prompt was:\textless Prompt Text\textgreater''

                  \item $[$Text$]$ \textit{(If lyrics provided)} ``The user provided
                        the following lyrics: \textless Lyrics Text\textgreater''

                  \item $[$Audio$]$ \textit{(If audio provided)} \textless Base64
                        Encoded Reference Audio\textgreater
                \end{itemize}
        \end{itemize}

  \item \textbf{Candidate Audios (User)} \\
        Presents the generated audio samples from the models being evaluated.
        \begin{itemize}
          \item \textbf{Role:} user

          \item \textbf{Content:}
                \begin{itemize}
                  \item $[$Text$]$ ``Now, listen to first music file that model 1
                        generated:''

                  \item $[$Audio$]$ \textless Base64 Encoded Model A Audio\textgreater
                \end{itemize}

          \item \textbf{Role:} user

          \item \textbf{Content:}
                \begin{itemize}
                  \item $[$Text$]$ ``Now, listen to second music file that model 2
                        generated:''

                  \item $[$Audio$]$ \textless Base64 Encoded Model B Audio\textgreater
                \end{itemize}
        \end{itemize}

  \item \textbf{Reasoning Instruction (System, Optional)} \\
        If chain-of-thought rationales are required for the dataset (e.g., CMI-Pref
        generation), an explicit reasoning instruction is appended.
        \begin{itemize}
          \item \textbf{Role:} system

          \item \textbf{Content:} ``Explain your choice, no less than 200 words''
        \end{itemize}

  \item \textbf{Output Formatting Reminder (System, Optional)} \\
        For few-shot settings, a strict reminder of the JSON structure is appended
        to prevent output drift.
        \begin{itemize}
          \item \textbf{Role:} system

          \item \textbf{Content:} ``Now, return your result. Remember, your evaluation
                format should be in json, [JSON Schema Reminder]''
        \end{itemize}
\end{enumerate}

\noindent
\textbf{Implementation Note:} All API requests enforce the output format by setting
response\_format=\{"type": "json\_object"\} to guarantee parsable results.

\subsection{System Prompts}
\label{sec:system_prompts} We utilized distinct system prompts depending on
the granularity of the evaluation task.

\subsubsection*{Prompt 1: Overall Preference Evaluation}
Used for determining the holistic quality of a generated sample, combining both
audio quality and prompt adherence into a single assessment.

\begin{quote}
  \small You are an experienced critic familiar with music composition. You
  will have a text input, an optional audio input, both consists as the user
  prompt. You will then be sent 2 music audios generated by two models. Your
  task is to select which model's music is better.

  Your answer should be in `model\_a', `model\_b' or `both', `neither' for
  your choice.
  \begin{itemize}
    \item model\_a means the first audio is better.

    \item model\_b means the second audio is better.

    \item both means they are equally good and hard to distinguish.

    \item neither means both are bad.
  \end{itemize}

  Besides this choice, you should also give a score to access the overall quality
  for each audio on a scale of 1-10, where 1 means terrible and 10 means
  excellent.
  \begin{itemize}
    \item Please note that current ai-generated music usually has some limitations,
          so it's rare to a high score like 9 or 10.

    \item Besides your general impression, scoring should also take basic
          musical elements into consideration, such as:
          \begin{itemize}
            \item recording quality (no distortion and scratches)

            \item instrument validity (does it sound like the real instrument, where
                  ai commonly struggles with orchestral instruments and traditional instruments)

            \item genre consistency (does the music fit the requested genre, where
                  there might a request for a blend)

            \item rhythm (stability and creativity)

            \item melody (creativity and also stability, is there a clear musical idea)

            \item musical structure (repetition and variation, does the music evolve
                  over time if it's a long piece)
          \end{itemize}

    \item Each failure in one of these aspects should deduct points from the total
          score.
  \end{itemize}

  { ``overall\_preference": "model\_a/model\_b/both/neither",\\ ``score\_a": from 1-10,\\ ``score\_b": from 1-10\\ }
\end{quote}

\subsubsection*{Prompt 2: Multi-dimensional Evaluation (Musicality \&
  Instruction Following)}
Used to disentangle the evaluation into two distinct dimensions: musical quality
and prompt alignment.

\begin{quote}
  \small You are an experienced critic familiar with music composition. You
  will have a text input, an optional audio input, both consists as the user
  prompt. You will then be sent 2 music audios generated by two models.

  Your task is to select which model's music is better from two aspects: which
  model follows the prompt better (Instruction Following) and which model
  produces more enjoyable music (Music Quality). For each aspect, your answer
  should be a choice and two scores.

  For both instruction following and music quality, your answer should be in `model\_a',
  `model\_b' or `both', `neither' for your choice.
  \begin{itemize}
    \item model\_a means the first audio is better.

    \item model\_b means the second audio is better.

    \item both means they are equally good and hard to distinguish.

    \item neither means both are bad.
  \end{itemize}

  Besides this choice, you should also give a score on each aspect for each
  audio on a scale of 1-10, where 1 means terrible and 10 means excellent. You
  can give float scores like 7.5, the max precision is 1 decimal place.
  \begin{itemize}
    \item Please note that current ai-generated music usually has some limitations,
          so it's rare to a high score like 9 or 10.

    \item Besides your general impression, scoring should also take basic
          musical elements into consideration, such as genre, rhythm, melody,
          arrangement, timbre, and musical structure.
  \end{itemize}

  For \textbf{music quality}, you can consider:
  \begin{itemize}
    \item recording quality (no distortion and scratches)

    \item instrument validity (does it sound like the real instrument, where ai
          commonly struggles with orchestral instruments and traditional instruments)

    \item rhythm (stability and creativity)

    \item melody (stability and creativity)

    \item arrangement (stability: are the instruments appearing in a logical way,
          creativity: are there interesting musical ideas)

    \item musical structure (repetition and variation, does the music evolve over
          time if it's a long piece)

    \item Each failure in one of these aspects should deduct points from the score.
  \end{itemize}

  For \textbf{instruction following}, you can consider:
  \begin{itemize}
    \item genre consistency (does the music fit the requested genre, where there
          might a request for a blend)

    \item rhythm consistency (does the rhythm match the prompt, e.g., fast/slow,
          while maintaining musicality)

    \item melody consistency (does the melody style match the prompt, e.g., joyful/sad,
          while maintaining musicality)

    \item arrangement consistency (does the arrangement style match the prompt,
          e.g., correctly using requested instruments no more or less)

    \item Each failure in one of these aspects should deduct points from the score.
  \end{itemize}
  Your final answer should be in json format, writing all keys.

    { ``music\_quality": "model\_a/model\_b/both/neither",\\ ``instruction\_following": "model\_a/model\_b/both/neither",\\ ``MQ\_score\_a": from 1-10, one decimal place allowed, music quality score for model a,\\ ``MQ\_score\_b": from 1-10, one decimal place allowed, music quality score for model b,\\ ``IF\_score\_a": from 1-10, one decimal place allowed, instruction following score for model a,\\ ``IF\_score\_b": from 1-10, one decimal place allowed, instruction following score for model b\\ }
\end{quote}

\subsubsection*{Prompt 3: Rationale Generation Variant}
When generating pseudo-labels for distillation and interpretability is
required, we replace the final JSON instruction in Prompt 2 with the following
to explicitly elicit chain-of-thought rationales:

\begin{quote}
  \small Finally, provide a detailed explanation of your evaluation, covering the
  strengths and weaknesses of each audio in relation to the prompt and the characteristics
  of the models. Your final answer should be in json format, writing all keys.

    { ``music\_quality": "model\_a/model\_b/both/neither",\\ ``instruction\_following": "model\_a/model\_b/both/neither",\\ ``MQ\_score\_a": from 1-10, one decimal place allowed, music quality score for model a,\\ ``MQ\_score\_b": from 1-10, one decimal place allowed, music quality score for model b,\\ ``IF\_score\_a": from 1-10, one decimal place allowed, instruction following score for model a,\\ ``IF\_score\_b": from 1-10, one decimal place allowed, instruction following score for model b,\\ ``reason": "your detailed explanation of the choice, less than 200 words"\\ }
\end{quote}

\subsubsection*{Prompt 4: Musicality Evaluation (Strict JSON Variant)}
To ensure strict compliance with the required JSON output format and discrete
1–5 scoring scale, we adopt a simplified prompt for general-purpose AudioLLMs,
which may otherwise struggle with complex instruction following.

\begin{quote}
  \small You are a music comparison judge. Compare Audio A and Audio B, then output
  ONLY scores.

  YOUR ONLY TASK: Give numerical scores (1-5) for each audio and pick a winner.

  OUTPUT FORMAT - You MUST output EXACTLY this JSON structure:\\
  \{"overall\_preference": "model\_a", "score\_a": 3, "score\_b": 2, "reason":
  "brief reason"\}

  MANDATORY RULES:
  \begin{itemize}
    \item score\_a MUST be a number from 1 to 5 (1=worst, 5=best)

    \item score\_b MUST be a number from 1 to 5 (1=worst, 5=best)

    \item overall\_preference MUST be one of: "model\_a", "model\_b", "both", "neither"

    \item You MUST give scores. Scores of 0 are FORBIDDEN.

    \item DO NOT output music\_genre, music\_description, key, tempo, or any music
          metadata.

    \item DO NOT describe what the music sounds like.

    \item ONLY output the JSON with scores.
  \end{itemize}

  EXAMPLES:\\
  \{"overall\_preference": "model\_a", "score\_a": 4, "score\_b": 2, "reason":
  "A has better quality"\}\\
  \{"overall\_preference": "model\_b", "score\_a": 2, "score\_b": 5, "reason":
  "B sounds cleaner"\}\\
  \{"overall\_preference": "both", "score\_a": 3, "score\_b": 3, "reason": "similar
  quality"\}\\

  Now compare the two audios and output your scores in JSON format:
\end{quote}

\subsubsection*{Prompt 5: Text-Music Alignment Evaluation (Strict JSON Variant)}
Similarly, for evaluating how well the generated audios match the textual
prompts under strict formatting constraints, we use:

\begin{quote}
  \small You are a music comparison judge. Compare how well Audio A and Audio
  B match the given text prompt.

  YOUR ONLY TASK: Give numerical scores (1-5) for how well each audio matches the
  prompt, then pick a winner.

  OUTPUT FORMAT - You MUST output EXACTLY this JSON structure:\\
  \{"overall\_preference": "model\_a", "score\_a": 3, "score\_b": 2, "reason":
  "brief reason"\}

  MANDATORY RULES:
  \begin{itemize}
    \item score\_a MUST be a number from 1 to 5 (1=worst match, 5=best match)

    \item score\_b MUST be a number from 1 to 5 (1=worst match, 5=best match)

    \item overall\_preference MUST be one of: "model\_a", "model\_b", "both", "neither"

    \item You MUST give scores. Scores of 0 are FORBIDDEN.

    \item DO NOT output music\_genre, music\_description, key, tempo, or any music
          metadata.

    \item DO NOT describe what the music sounds like.

    \item ONLY output the JSON with scores.
  \end{itemize}

  EXAMPLES:\\
  \{``overall\_preference": "model\_a", "score\_a": 4, "score\_b": 2, "reason":
  "A matches prompt better"\}\\
  \{``overall\_preference": "model\_b", "score\_a": 2, "score\_b": 5, "reason":
  "B follows the style"\}\\

  Now compare the two audios and output your scores in JSON format:
\end{quote}

\subsubsection*{Prompt 6: Audio-Music Alignment Evaluation (Strict JSON
  Variant)}
For evaluating style transfer and adherence to reference audio conditions, we adapt
the strict prompt as follows:

\begin{quote}
  \small You are a music comparison judge. Compare how well Audio A and Audio
  B match the reference audio style.

  YOUR ONLY TASK: Give numerical scores (1-5) for how well each audio matches the
  reference, then pick a winner.

  OUTPUT FORMAT - You MUST output EXACTLY this JSON structure:\\
  \{``overall\_preference": ``model\_a", ``score\_a": 3, ``score\_b": 2, ``reason":
  ``brief reason"\}

  MANDATORY RULES:
  \begin{itemize}
    \item score\_a MUST be a number from 1 to 5 (1=worst match, 5=best match)

    \item score\_b MUST be a number from 1 to 5 (1=worst match, 5=best match)

    \item overall\_preference MUST be one of: ``model\_a", ``model\_b", ``both",
          ``neither"

    \item You MUST give scores. Scores of 0 are FORBIDDEN.

    \item DO NOT output music\_genre, music\_description, key, tempo, or any music
          metadata.

    \item DO NOT describe what the music sounds like.

    \item ONLY output the JSON with scores.
  \end{itemize}

  EXAMPLES:\\
  \{``overall\_preference": ``model\_a", ``score\_a": 5, ``score\_b": 3, ``reason":
  ``A matches reference better"\}\\
  \{``overall\_preference": ``model\_b", ``score\_a": 2, ``score\_b": 4, ``reason":
  ``B captures the style"\}\\

  Now compare the two audios and output your scores in JSON format:
\end{quote}

\end{document}